\documentclass{JHEP3}

\usepackage{amssymb,amsmath} 
\usepackage{amsfonts}
\usepackage{graphicx}
\usepackage{epic}
\usepackage{eepic}
\usepackage{epsf}
\def\Xint#1{\mathchoice
   {\XXint\displaystyle\textstyle{#1}}%
   {\XXint\textstyle\scriptstyle{#1}}%
   {\XXint\scriptstyle\scriptscriptstyle{#1}}%
   {\XXint\scriptscriptstyle\scriptscriptstyle{#1}}%
   \!\int}
\def\XXint#1#2#3{{\setbox0=\hbox{$#1{#2#3}{\int}$}
     \vcenter{\hbox{$#2#3$}}\kern-.5\wd0}}
\def\dashint{\Xint\diagup}

\newcommand{\tF}{\tilde{F}} \newcommand{\tI}{\tilde{I}}

\title{\vspace{3cm} Descent Relations in Cubic Superstring Field Theory}

\author{\vspace{1cm} \text{I.Ya.~Aref'eva${}^1$, R.~Gorbachev${}^{1}$,
P.B.~Medvedev${}^{2}$, D.V.~Rychkov${}^{3,4}$}\\ \\
\textit{\small ${}^1$ Steklov Mathematical Institute of Russian
Academy of
Sciences,}\\ \textit{\small Gubkin st. 8, 119991, Moscow, Russia,}\\
\texttt{\small arefeva@mi.ras.ru, rgorbachev@mi.ras.ru}
\\ \textit{\small ${}^2$ Institute of Theoretical and Experimental Physics,}\\
\textit{\small B.Cheremushkinskaya st. 25, Moscow, 117218,}\\
\texttt{\small pmedvedev@itep.ru}
\\ \textit{\small ${}^3$ Physics Department, Moscow State
University,}\\ \textit{\small Moscow, Russia, 119899,}\\
\textit{\small ${}^4$ Department of Physics and Astronomy,
University of Southern
California,}\\ \textit{\small Los Angeles, CA 90089-0484,}\\
\texttt{\small rychkov@usc.edu}}

\abstract{ The descent relations between  string field theory (SFT) vertices are
characteristic relations of the operator formulation of SFT and they provide
self-consistency of this theory. The descent relations $\langle V_2|V_1\rangle$ and
$\langle V_3|V_1\rangle$ in the $NS$ fermionic string field theory in the $\kappa$
and discrete bases are established. Different regularizations and schemes of
calculations are considered and relations between them are discussed. }

 \preprint{arXiv:0704.3688}

\begin{document}

\section{Introduction}

The cubic open superstring field theory (SSFT) which keeps the main features of the
Witten open bosonic string field theory \cite{witten} has the following action
\cite{AMZ,PTY}:
\begin{equation} S= \dashint\Phi\star Q\Phi + \frac23\dashint\Phi
\star \Phi\star \Phi. \label{w}
\end{equation}
As in the bosonic SFT action the
action (\ref{w}) contains two main ingredients: a multiplication $\star$ and an
integral $\dashint$. Comparing with its bosonic counterpart $\int$, $\dashint$ has an
extra picture changing operator: $\dashint\equiv\int Y_{-2}$ \cite{AMZ,PTY}.

An appearance of the notion of pictures is a main novelty in the
open SSFT. The pictures are the characteristics of the ghost sector
of the fermionic string. We explain all specific of the appearance
of picture changing operators in Section 2.

Representations of $\star$ and $\dashint$ in the operator and
conformal languages were constructed in \cite{GJ, LPP, Vertex} (for
a review see \cite{Ohmori, ABGKM, Taylor:2003gn,LB}). Initially the
vertices in the operator realization (both for the bosonic and for
the fermionic strings) were built in the discrete basis (standard
oscillator basis) and defined through the infinite dimensional
Neumann matrices \cite{GJ, Vertex}. Calculations using these
vertices are rather complicated \cite{Taylor}. Rastelli, Sen and
Zwiebach \cite{RSZ} suggested to transform  vertices to the basis
called $\kappa$-basis where $K_1 =L_1+L_{-1}$ is diagonal. The
formalism of the $\kappa$-basis has already proved its simplicity in
\cite{Belov}-\cite{Bonora:2007tm}.

In the operator representation of the theory (\ref{w}) fields $\Phi$
are realized as ket-vectors $|\Phi\rangle$ in the Fock space. The
multiplication and the integral can be given with the help of two
vertices. The vertex $\langle |\hat{V}_3 |\rangle$  represents
$\star$:
$$
|\Phi_1\star\Phi_2 \rangle=\langle |\hat{V}_3 |\rangle
|\Phi_1\rangle |\Phi_2\rangle
$$
and the vertex $\langle \hat{V}_1|$  represents $\dashint$:
$$
\dashint \Phi=\langle \hat V_1 |\Phi_1\rangle.
$$
Having these vertices one can construct pure "left" vertices
$\langle \hat{V}_2|,$  $\langle \hat{V}_3|,$ $\langle \hat{V}_4|,$... etc. via
natural definitions
$$
\langle \hat{V}_2|=\langle \hat{V}_1|\langle |\hat{V}_3 |\rangle , \,\,
\langle \hat{V}_3|=\langle \hat{V}_1|\langle |\hat{V}_3 |\rangle
\langle |\hat{V}_3 |\rangle ,\,\,
\langle \hat{V}_4|= \langle \hat{V}_1|\langle |\hat{V}_3 |\rangle
\langle |\hat{V}_3 |\rangle \langle|\hat{V}_3 |\rangle ,...
$$
These vertices form the so-called Witten tower of vertices
\cite{AGMR}. They  satisfy the descent relations which are written
in terms of the ``right" 1-vertex $|\hat{V}_1\rangle$
 defined as a solution of the following relation
 \begin{equation}
 \langle\hat{V}_2||\hat{V}_1\rangle=\langle\hat{V}_1| .
 \label{**}
 \end{equation}
One can prove \cite{AGMR} that the descent relations have the form
\begin{equation}
\langle\hat{V}_{n+1}||\hat{V}_1\rangle=\langle\hat{V}_n| .
\label{*}
\end{equation}

 In practice one usually construct the "left" vertices
 $\langle V_{n}|$ by solving  overlap conditions \cite{GJ}.
 To stress the origin of these vertices we remove hats.
 It is obvious that the overlap conditions define vertices up to normalization
 factors
 $\langle\hat{V}_n|=Z_n \langle V_n|$ and $|\hat{V}_{1}\rangle =Z_{-1} | V_1
 \rangle$.
Therefore  the following modification
 of the descend relations (\ref{*}) for $\langle V_n | $ takes place
 \begin{equation}
 \langle V_{n+1}\mid V_1\rangle=\mathcal{Z}_n \langle V_n|  ~~~~~
 \mathcal{Z}_n =Z_nZ_{-1}^{-1}Z_{n+1}^{-1}
 \label{2.9}
\end{equation}

 These relations are indispensable ingredients of SFT,
 and they play important role in the SFT perturbation technique \cite{Taylor}.
 Note that the vertices $\hat{V}_1$ and $\hat{V}_3$ are defined up to
 two normalization factors as well. These factors can be absorbed
 into the charge and string field redefinitions in the cubic SFT action.

The descent relations are exact relations.  Performing a check of
the descent relations within a calculation scheme one performs a
check of the calculation scheme itself. One can analyze
\begin{itemize}
    \item operator structure of relations (\ref{2.9});
    \item  numerical factors in (\ref{2.9}).
\end{itemize}
 There is no reason to expect that $\mathcal{Z}_n$ are well defined
 and therefore a regularization is needed. Different regularizations are
 suitable for different bases and one can expect different factors
 $\mathcal{Z}_n$ for different schemes and different
 regularizations. Moreover the $\mathcal{Z}_n$ is a product of factors coming
 from matter and ghost sectors separately. Each of these factors
  are singular ones tend to zero and another ones to infinity. One needs
  a special relation between matter and ghost regularizations to provide finite
  answers.

 A study of the descent relations between string vertices in the
bosonic string has been performed by Belov \cite{Belov2} and Fucks
and Kroyter \cite{KF}. In ref.
\cite{Belov2,KF,AGMR,KF2} it was found that $\mathcal{Z}_3\neq1$.
The appearance of $\mathcal{Z}_3\neq1$ in the descent relations was
considered as an anomaly \cite{KF}. In  \cite{KF2} there
was suggested the regularization for which the coefficient
$\mathcal{Z}_3$ was equal one in the bosonic string.
The factorization of the coefficient
$\mathcal{Z}_n=Z_n Z_{-1}^{-1}Z_{n+1}^{-1}$ was found in the
$\kappa$-basis \cite{Belov2}, the origin of this factorization was
revealed in \cite{AGMR}.

In this paper we investigate the descent relations  in the
Neveu-Schwartz $(NS)$ sector of the fermionic string. Our goals are
\begin{itemize}
    \item find a numerical factor in the relation $\langle V_2|
V_1\rangle=\mathcal{Z}_2\langle V_1|$;
    \item check the descent relation $\langle V_3| V_1\rangle = \mathcal{Z}_3\langle
    V_2|$.
\end{itemize}

 We check these relations  in the $\kappa$-basis and in the discrete one and
discuss the different regularizations in these bases. It is
interesting to note that in contrast to the calculations in the
bosonic string where all calculations in the discrete  basis were
performed only numerically \cite{KF,AGMR}, some calculations in the
fermionic sector can be performed analytically. Namely, we analytically
calculate the Neumann matrix in the vertex $\langle V_2|$ through a
multiplication the Neumann matrices of the vertices $\langle V_3|$
and $|V_1\rangle$.

This paper is organized as the following. In Section 2 we remind
some aspects of the Fock representation of the superstring field
theory and note special aspects of the descent relations for the NS
string. In Section 3 we discuss some properties of the lowest
vertices $V_1$ and $V_2$. In Section 4 we remind some features of
the $\kappa$-basis for the $NS$ fermionic string and calculate the
descent relation between vertices $\langle V_3|$ and $| V_1\rangle$.
Here we also calculate the coefficient $\mathcal{Z}_3$. The descent
relation $\langle V_3|V_1\rangle = \langle V_2| $ in the discrete
basis is calculated in Section 5. In Section 6 we discuss the
coefficients $\mathcal{Z}_3$ and an issue of  regularizations.
\section{Specific of Decent Relations in NS string}

Let us remind the main features of the fermionic string. The ghost
$\beta,~\gamma$ sector  of the fermionic string has many different
nonequivalent vacua $|q\rangle$ which are known as pictures and
labeled by integers. Transitions between vacua (pictures) are
realized by an infinite number of ghost modes. This is the
difference between the ghost sector of the fermionic string and the
ghost sector of the bosonic string where the transition between
vacua is realized by one mode of ghost or anti-ghost. There is an
operator which changes the picture:
\begin{equation}
|q\rangle=e^{q\phi(0)}|0\rangle,
\end{equation} where $\phi(z)$ is the field
bozonizing the superghosts as
$\beta=e^{-\phi}\partial\xi,~\gamma=e^{\phi}\eta$ \cite{FMS}. The
BRST-invariant operators realizing the transition between various
vacua are called the picture changing operators
\cite{AMZ,PTY,FMS,AM}. There are two picture changing operators $X$
and $Y_{-1}$ which change the picture $|q\rangle$ in the following
way
\begin{eqnarray}
X(|q\rangle)&\Rightarrow &|q+1\rangle,\nonumber\\
Y_{-1}(|q\rangle)&\Rightarrow &|q-1\rangle.
\end{eqnarray}
Both these operators commute with BRST charge $Q$: $
[X,Q]=0,\quad[Y_{-1},Q]=0, $ and obey the key property $ (X\cdot
Y_{-1})(z)=\lim_{z\to z'}X(z)Y_{-1}(z')=1$.

The normalization of the vacua has the form \cite{FMS}
\begin{eqnarray}
\label{vacua-norm}
\langle -q-2|q\rangle=1.
\end{eqnarray}
In the action \eqref{w}
one uses zero-picture string fields
\begin{equation} |\Phi^{(0)}\rangle=\sum _{k,l,n
\in \mathbb{N} \atop m,j,i \in \mathbb{N} +\frac12}A_{i\dots}(x)\,\beta_{-i}
...\gamma_{-j}...b_{-k}...c_{-l}...\alpha^{\mu}_{-n}... \psi^{\nu}_{-m} |0\rangle.
\label{f}
\end{equation}
The vertex $\langle \hat{V}_1|$ defining the string integral for zero picture fields
is a vector and has the picture ``-2'' and we denote it by the
superscript  ``(-2)''
\begin{equation}
\label{int} \dashint \Phi\equiv
^{(-2)}\langle \hat{V}_1|\Phi^{(0)}\rangle.
\end{equation}

Next one defines the star product $\star$ of two string fields as
\begin{equation}
\label{star} |\Phi_1\star\Phi_2\rangle_3^{(0)}={}_{12}\langle
|\hat{V}_3|\rangle_3|\Phi^{(0)}\rangle_1|\Phi^{(0)}\rangle_2.
\end{equation}
Hence the
vertex ${}_{12}\langle |\hat{V}_3|\rangle_3$ in (\ref{star}) has the pictures ``-2'',
``-2'' and ``0'' respectively, so we denote it as ${}^{(-2,-2)}{}_{12}\langle
|V_3|\rangle_3^{(0)}$.

Using the vertices ${}^{(-2)}\langle \hat{V}_1|$ and ${}^{(-2,-2)}{}_{12}\langle
|\hat{V}_3|\rangle_3^{(0)}$ one can build a vertex $\langle \hat{V}_2|$ as
\begin{equation}
\label{V_2_def}
{}^{(-2,-2)}{}_{12}\langle \hat{V}_2|\equiv{}^{(-2)}{}_3
\langle
\hat{V}_1|^{(-2,-2)}{}_{12}\langle |\hat{V}_3|\rangle_3^{(0)}.
\end{equation}
Now we are able to define a vertex $| \hat{V}_1\rangle$ as a
solution of eq. (\ref{**}) with $|\hat{V}_1 \rangle$ in the zero
picture:
\begin{equation}
\label{2} {}^{(-2,-2)}{}_{12}\langle \hat{V}_2|
\hat{V}_1\rangle_2^{(0)}\equiv{}^{(-2)}{}_1\langle \hat{V}_1|.
\end{equation}

 One can build an infinite tower of vertices ${}_{1\dots
N}\bigl\langle|\hat{V}_{N+1}|\bigr\rangle_{N+1}$ by gluing of $N-1$ vertices~
$\hat{V}_3$. They
have $N$ vacua in the picture ``-2'' and one vacuum in the picture ``0''. Using the
vertex ${}_{1\dots N}\bigl\langle|\hat{V}_{N+1}|\bigr\rangle_{N+1}$, one can define the
star product for~$N$ fields as:
\begin{equation}
{}^{(-2,-2,...-2)}{}_{1\dots
N}\bigl\langle|\hat{V}_{N+1}|\bigr\rangle^{(0)}_{N+1}
|\Phi^{(0)}_N\rangle\dots |\Phi^{(0)}_1\rangle\equiv
|\Phi_N^{(0)}\rangle\star\dots\star|\Phi_1^{(0)}\rangle.
\end{equation}

 By an analogy with the bosonic case one can also build an infinite tower
 of the
vertices ${}^{(-2,...,-2)}{}_{1\dots N}\langle \hat{V}_{N}|$
associated with the vertices ${}^{(-2,...-2)}{}_{1\dots
N}\langle|\hat{V}_{N+1}|\rangle_{N+1}^{(0)}$ constructed above by
adding one more ${}^{(-2)}\langle \hat{V}_1|$ as in (\ref{V_2_def}).
Like we did it \cite{AGMR} in the bosonic string, let us call this
set of the vertices Witten's tower of vertices.  Therefore one has
the
  descent relations between the vertices $\langle \hat{V}_{N+1}|$ and $\langle
  \hat{V}_{N}|$ \cite{AGMR}:
\begin{equation}
{}^{(-2,...,-2)}\langle \hat{V}_{N+1}|\hat{V}_1\rangle^{(0)}=
{}^{(-2,...,-2)}\langle \hat{V}_N|.
\end{equation}

As has been mentioned in the introduction in practice one defines
$\langle V_{N}|$ via the overlap conditions that define
corresponding vertices up to numerical factors. Moreover, in
practice defining a vertex via these overlap conditions one can use
different realizations of   pictures. For example,
${}^{(-2,-2)}\langle V_{2}|$ can be presented with help of an
operator acting
 on a bra vacuum vector $\langle -2|\langle -2 |$
 \begin{equation}
 \label{V2-1}{}^{(-2,-2)}\langle V_{2}|=\langle -2|{}
 \langle -2|\,\,{\cal V}_{-2,-2} ,
 \end{equation}
as well as result of  acting of the picture changing operator
$Y_{-1}Y_{-1}$ (in different points) on a vector $\langle -1|\langle
-1|\,\,{\cal V}_{-1,-1}$,
\begin{equation}
 \label{V2-2}{}^{(-2,-2)}\langle V_{2}|=\langle -1|\langle -1|\,\,
 {\cal V}_{-1,-1}\,\,Y_{-1}Y_{-1},
 \end{equation}
or
\begin{equation}
 \label{V2-3}{}^{(-2,-2)}\langle V_{2}|=\langle 0|\langle -2|\,\,
 {\cal V}_{0,-2}\,\,Y_{-2},
 \end{equation}
with $Y_{-2}$ being $Y_{-2}|q\rangle\Rightarrow|q-2\rangle$.

It is difficult to treat $Y_{-1}$ in operator formalisms  \cite{AM}
and therefore, it is difficult to check explicitly that these
representations (\ref{V2-1}), (\ref{V2-2}) and (\ref{V2-3}) are
equivalent (see also \cite{T}). As for the descent relation
 $
{}^{(-2,-2,-2)}\langle
 V_3|V_1\rangle^{(0)}\simeq {}^{(-2,-2)}\langle V_2|
 $
we will check it for vertices in the form $ {}^{(-2,-2,-2)}\langle
V_3|=
\langle-1|\langle-1|\langle-2|\mathcal{V}_{-1,-1,-2}Y_{-1}Y_{-1} $
and $ {}^{(-2,-2)}\langle V_2|
=\langle-1|\langle-1|\mathcal{V}_{-1,-1}Y_{-1}Y_{-1}, $ i.e. we have to
check

\begin{equation}
\label{10}
 \langle-1|\langle-1|\langle-2|\mathcal{V}_{-1,-1,-2}Y_{-1}Y_{-1}|V_1\rangle=
\langle-1|\langle-1|\mathcal{V}_{-1,-1}Y_{-1}Y_{-1} .
\end{equation}

We can assume that $Y_{-1}$-'s acts on one ``external" legs 1, or 2
in (\ref{10}) and remove $Y_{-1}Y_{-1}$ that gives
\begin{equation}
{}^{(-1,-1,-2)}\langle V_3|V_1\rangle^{(0)}\simeq{}^{(-1,-1)}\langle
V_2|.
\end{equation}
The same is true for $|V_1\rangle$ defining relation. We actually
check
\begin{equation*}
{}^{(-2,0)}\langle V_2|V_1\rangle^{(0)}={}^{(0)}\langle V_1|
\end{equation*}
as
\begin{equation*}
{}^{(-2,-2)}\langle V_2|V_1\rangle^{(0)}={}^{(-2)}\langle V_1|
\end{equation*}
in a special form is equivalent to
\begin{equation*}
\langle-2,0|\mathcal{V}_{-2,0}Y_{-2}|V_1\rangle=\langle0|\mathcal{V}_0Y_{-2}
\end{equation*}
  Therefore for checking the descent relations
we have a freedom to  choose the vertices in the pictures which are more
convenient for the calculations.

\section{$V_1$ and $V_2$}
The solution $| V_1\rangle$ of the defining equation \eqref{2} is
unique, that can be transparently exemplified on the bosonic vertex
$| V_1^X\rangle$. To show this it is enough to check that $\langle
V_2|$ is nondegenerated. In the case of bosonic string
\begin{equation}
\langle V_2^X|={}_{1,2}\langle 0|\exp(-a^1Ca^2),
\end{equation} where
$[a_n,a^{\dag}_m]=\delta_{nm}$ are the modes of the string $X(\sigma)$ and
$C_{nm}=(-1)^n\delta_{nm}$ (here we omit the zero mode for simplicity). Reminding the
correspondence between a string field $\Phi[X(\sigma)]$ and a state $|\Phi\rangle$:
\begin{equation}
\Phi[X(\sigma)]\equiv\langle X(\sigma)|\Phi\rangle
\end{equation}
we write the Dirac conjugation
\begin{equation}\label{Dirac} \overline{\Phi}
[X(\sigma)]=\langle \Phi|X(\sigma)\rangle.
\end{equation}
Considering the vertex
$\langle V_2^X|$ as an operator mapping ket-s $|\Phi\rangle$ to bra-s $\langle
\Phi^R|$\footnote{To avoid misunderstanding we stress that string fields are
represented as ket-s: $\Phi[X(\sigma)]\equiv \langle X(\sigma)|\Phi\rangle$ not as
bra-s: $\langle \Phi|X(\sigma)\rangle=\overline{\langle
X(\sigma)|\Phi\rangle}=\overline{\Phi}[X(\sigma)]$.}
\begin{equation}
\label{Phi-R}
{}_2\langle \Phi^R|\equiv {}_{1,2}\langle V_2|\Phi\rangle_1
\end{equation} we write
using (\ref{Dirac}):
\begin{equation}
\label{phi-tilde} \overline{\Phi}^R
[X(\sigma)]={}_{1,2}\langle V_2|\Phi\rangle_1|X(\sigma)\rangle_2.
\end{equation} We
calculate (for $E^{-1}_{nm}$ see \cite{GJ})
\begin{eqnarray}
{}_{1,2}\langle
V_2|X(\sigma)\rangle_2&=&{}_{1,2}\langle 0| \exp(-a_n^1(-1)^n
a_n^2)\exp\left(-x_nE^{-2}_{nn}x_n-2ia^{1\dag}_nE^{-1}_{nn}x_n+\frac 1 2
a^{1\dag}_na^{1\dag}_n\right)|0\rangle_1\nonumber\\ &=&{}_{2}\langle 0|
\exp\left(-x_nE^{-2}_{nn}x_n+2i(-1)^n a^2_nE^{-1}_{nn}x_n+\frac 1 2
(-1)^na^2_n(-1)^na^2_n\right)\\ &=&{}_{2}\langle 0|
\exp\left(-((-1)^nx_nE^{-2}_{nn}(-1)^nx_n)+2i a^2_nE^{-1}_{nn}(-1)^nx_n+\frac 1 2
a^2_na^2_n\right).\nonumber
\end{eqnarray}
Taking into account the string modes
expansion $X(\sigma)=i\sqrt{2}\sum_{n}x_n\cos(n\sigma)$ one gets $${}_{1,2}\langle
V_2|X(\sigma)\rangle_2=\langle X(\pi-\sigma)|.$$ That gives
\begin{equation}
\overline{\Phi}^R [X(\sigma)]=\langle X(\pi-\sigma)|\Phi\rangle=\Phi[X(\pi-\sigma)].
\end{equation}
There is a string field decomposition
$$\Phi[X(\sigma)]=\Phi_S[X(\sigma)]+\Phi_A[X(\sigma)]$$ into the symmetric part
$$\Phi_S[X(\sigma)]=\Phi_S[X(\pi-\sigma)]$$ and the antisymmetric part
$$\Phi_A[X(\sigma)]=-\Phi_A[X(\pi-\sigma)].$$ Under this decomposition we have
\begin{equation}
\Phi^R[X(\sigma)]=\overline{\Phi}_S[X(\sigma)]-\overline{\Phi}_A[X(\sigma)].
\end{equation}
Hence, we see that $\langle V_2|$ acts as the mapping
\begin{equation}
\langle V_2|:\quad \Phi[X(\sigma)]=\Phi_S[X(\sigma)]+\Phi_A[X(\sigma)]\to
\overline{\Phi}_S[X(\sigma)]-\overline{\Phi}_A[X(\sigma)].
\end{equation}
This
mapping is nondegenerate and by this reason the solution of (\ref{2}) is unique.

Taking $\langle V_2|$ as a quadratic form in the space of the string functional
$(\Phi,\Phi)\equiv {}_{1,2}\langle V_2|\Phi\rangle_1|\Phi\rangle_2$ one sees that
this metrics is neither positive defined nor nondegenerate. Indeed,
\begin{eqnarray}
\label{VFF} \langle V_2|\Phi\rangle|\Phi\rangle &=&\langle
\Phi^R|\Phi\rangle=\langle
\overline{\Phi}_S-\overline{\Phi}_A|\Phi_S+\Phi_A\rangle=\nonumber\\ &=&\langle
\overline{\Phi}_S|\Phi_S\rangle-\langle \overline{\Phi}_A|\Phi_A\rangle+\langle
\overline{\Phi}_S|\Phi_A\rangle-\langle \overline{\Phi}_A|\Phi_S\rangle=\nonumber\\
&=&\langle \overline{\Phi}_S|\Phi_S\rangle-\langle \overline{\Phi}_A|\Phi_A\rangle.
\end{eqnarray}

A few comments concerning $| V_1\rangle$ are in order. Let us
multiply both sides of \eqref{2} by an arbitrary field
$|\Phi\rangle$ and rewrite it in the Witten notations
\begin{equation}
\int V_1\star \Phi=\int \Phi,
\end{equation}
where $ V_1$ stand for a string field corresponding to $|
V_1\rangle$ by $V_1[X(\sigma)]\equiv\langle X(\sigma)| V_1\rangle$.
Since $\Phi$ is an arbitrary field one can discard the integral to
get
\begin{equation}
 V_1\star \Phi=\Phi.
\end{equation}
 Thus we conclude that $V_1$ defined by
\eqref{2} is nothing but the unity under the star multiplication
\cite{GJ,RZ,Ellwood,KO}.

Multiplying both sides of \eqref{Phi-R} by $| V_1\rangle$ from the
right one gets
\begin{equation}
 \langle V_2|\Phi\rangle |
V_1\rangle=\langle \Phi^R| V_1\rangle,
\end{equation}
that gives via
descent relation \eqref{2}
\begin{equation}
\label{phi-equat}
\langle V_1|\Phi\rangle=\langle\Phi^R| V_1\rangle.
\end{equation}
In order to have the positively defined and nondegenerate
quadratic form (\ref{VFF}) we put the following constraints on the symmetric and
antisymmetric field parts:
\begin{eqnarray}
\Phi_S=\overline{\Phi}_S,&\text{i.e}&\text{the symmetric part is a real
one,}\nonumber\\
 \Phi_A=-\overline{\Phi}_A,&\text{i.e}&\text{the antisymmetric part is an imaginary
 one,}
\end{eqnarray}
or for the field $\Phi$ we get
\begin{equation}\label{real}
\Phi[X(\sigma)]=\Phi^R[X(\sigma)].
\end{equation}
 Following \cite{witten,Samuel} we
call these fields as real. We see that on the linear subspace of the
real string field $\langle V_2^X|$ acts as a unity operator i.e. $
\langle V_2^X | \Phi \rangle = (|\Phi\rangle )^+ $. The interesting
question arises if the subspace of the real field is a subspace
under $\star$ multiplication.

Hence, turning back  to (\ref{phi-equat}) we have for the real fields
\begin{equation}
\label{dirac-conj}
 \langle V_1|\Phi\rangle=\langle\Phi|V_1\rangle.
\end{equation}
Thus, we conclude that $| V_1\rangle$ is both
the identity for
 $\star$ and the ket representation of the Witten integral for the
real string fields (\ref{real}).

We start our calculation with the defining relation (\ref{2}) for the fermionic
variables $\psi$. By the direct calculation we check (\ref{2}) for given vertices
\cite{GJ,T}.

The vertex $|V_1^\psi\rangle$ has the form \cite{GJ}
\begin{equation}
|V_1^\psi\rangle=\exp\left\{\frac12\sum_{r,s\geq1/2}\psi_{-r}^\mu
I_{rs}\psi_{-s}^\mu\right\}|0\rangle=\exp\left(\frac 1 2 \psi
I\psi\right)|0\rangle,
\end{equation} where $I_{rs}$ is the antisymmetric matrix (see Appendix, eq.
(\ref{V-matrices})).

The vertex $\langle V_2^{\psi}|$ reads \cite{GJ}
 \begin{eqnarray}
 \langle
V_2^{\psi}|={}_{12}\langle 0|\exp\left(-i\sum_{r\geq 1/2}\psi^1_r (-)^r
\psi^2_r\right)={}_{12}\langle 0|\exp(-\psi^1 S\psi^2),
\end{eqnarray}
where
$S_{rs}=i(-1)^r\delta_{r,s}$.

So we have
\begin{eqnarray}
\langle
V_2^{\psi}|V_1^{\psi}\rangle={}_{12}\langle 0|
\exp(-\psi^1S\psi^2)\exp(\frac 1 2 \psi^1 I\psi^1)|0\rangle_{1}.
\end{eqnarray}
We use the following identity
\cite{AGM}
\begin{eqnarray}
\label{ff}
\langle0|\exp\left\{\frac12\psi_rV_{rs}\psi_s+\psi_r\lambda_r\right\}
\exp\left\{\frac12\psi_{-r}I_{rs}\psi_{-s}\right\}|0\rangle\nonumber\\
=\det(1+VI)^{1/2}\exp\left\{\frac12\lambda_r(1+IV)^{-1}I_{rs}\lambda_s\right\}.
\end{eqnarray}
We get
\begin{eqnarray}
\langle
V_2^{\psi}|V_1^{\psi}\rangle=\langle 0|\exp\left(-\frac 1 2
\psi_r^2(-1)^{r+p}I_{rp}\psi^2_p\right)=\langle V_1^{\psi}|.
\end{eqnarray}
 For the superghost sector we can also calculate
$\langle V_2|V_1\rangle$. The vertex $|V_1^{\beta\gamma}\rangle$
reads \cite{GJ}.
\begin{equation}
|V_1^{\beta\gamma}\rangle=\exp\left(\sum_{r\geq 3/2 \atop s\geq
-1/2} \beta_{-r}K_{rs}\gamma_{-s}\right)|0\rangle.
\end{equation}
The vertex $\langle V_2|$ has the following form
\cite{T}
\begin{equation}
 \label{V2}\langle V_2^{\beta\gamma}| ={}_{1,2}\langle -2,0| \exp\left(i\sum_{r\geq
 -1/2}
(-1)^r \gamma_r^2\beta_r^1-i\sum_{r\geq 3/2} (-1)^r
\gamma_r^1\beta_r^2\right).
\end{equation}

The descent relation is the following
\begin{eqnarray}
 \langle
V_2^{\beta\gamma}|V _1^{\beta\gamma}\rangle={}_{1,2}\langle
-2,0|\exp\left(i\sum_{r\geq -1/2}(-1)^{r}\gamma_r^2 \beta_{r}^1
-i\sum_{r\geq
3/2}(-1)^{r}\gamma_r^1 \beta_{r}^2\right)\times\nonumber\\
\times\exp\left(\sum_{r\geq 3/2 \atop s\geq -1/2}
\beta_{-r}^1K_{rs}\gamma_{-s}^1\right)|0\rangle _1.
 \end{eqnarray}
  We use the following
identity \cite{AGK}
\begin{eqnarray}
\label{gg}
\langle-2|\exp\{-\beta_rS_{rs}\gamma_s+\lambda_s\gamma_s+\beta_s\mu_s\}
\exp\{\beta_{-r}V_{rs}\gamma_{-s}+\beta_{-s}\nu_{s}+\kappa_s\gamma_{-s}\}|0\rangle\\
=\det(1-SV)^{-1}
\exp\{\kappa(1-SV)^{-1}(S\nu-\mu)-\lambda(1-VS)^{-1}(V\mu-\nu)\}\nonumber.
\end{eqnarray}
 So we get
\begin{equation} \langle
V_2^{\beta\gamma}|V_1^{\beta\gamma}\rangle=\langle
0|\exp\left(-\sum_{r\geq3/2\atop s\geq-1/2} \beta_r^2
(-1)^{r+s}K_{rs}\gamma_s^2\right)=\langle V_1^{\beta\gamma}|.
\end{equation}
Here we got that the vertices $|V_1^\psi\rangle$ and
$|V_1^{\beta\gamma}\rangle$ satisfy the defining equation (\ref{2})
(the vertices $|V_1^\psi\rangle$ and $|V_1^{\beta\gamma}\rangle$
were got as the solution of the overlap equation). We also got the
coefficient $\mathcal{Z}_2$
equal to one.
\section{Descent relation in the $\kappa$-basis }
\subsection{$\kappa$-basis. General}
The excellent description of the $\kappa$-basis for the fields of an arbitrary
conformal weight $s$ was given in \cite{BL}. For the $NS$ sector of superstring we
will be interested in the case $s=1/2$. In this section we review the notations and
the main formulas from \cite{BL,AG}.

Let $\mathcal{H}_s$ be the Hilbert space of analytic functions inside the unit disk
and square-integrable on the boundary. The inner product for $f,g\in\mathcal{H}_s$ is
given by
\begin{equation}
 \langle
g|f\rangle_s=\frac{1}{\pi\Gamma(2s-1)}\,\int_{|z|\leqslant 1} d^2z\,\bigl[1-z\bar
z]^{2s-2}\, \overline{g(z)}f(z). \label{norm}
 \end{equation}
  In our case $s=1/2$. The
apparent singularity at $s=\frac12$ is spurious \cite{ruhl}.

The usual oscillator basis diagonalizes the generator $L_0$, which has discrete
eigenvalues $(m+s)$, $m=0,1,2,\dots$. Its eigenfunctions normalized by \eqref{norm}
are
\begin{equation}
|m,s\rangle(z)= N_m^{(s)} z^m \quad\text{with}\quad
N_m^{(s)}=\left[\frac{\Gamma(m+2s)}{\Gamma(m+1)}\right]^{1/2}. \label{basisn}
\end{equation}
 For $s=1/2$ the normalization factor $ N_m^{(1/2)}$ equals to one and
$|m,1/2\rangle(z)=z^m$.

The vertices $\langle V_N|$ are $K_n$ symmetrical ones
 \begin{eqnarray}
\label{K_n-sym} \langle V_3|(K^{(1)}_n+K^{(2)}_n+K^{(3)}_n)&=&0,
\end{eqnarray}
where
$K_n=L_n-(-)^nL_{-n}$ and $L_n$ are the Virasoro generators. It is important that for
$n=even$ only the full (matter+ghost) vertex (\ref{K_n-sym}) is $K_n$-invariant, but
for $n=odd$ the vertices are invariant in the matter and ghost sectors \cite{GJ,RSZ}
separately. For $n-odd$ the $K_n$ invariance of a vertex means that the Neumann
matrices $V_{kl}$ and the matrix corresponding to the operator $K_n$ commute (choose
$n=1$):
\begin{equation}
 [K_1,V]=0.
 \end{equation}
  Therefore, if one finds the
eigenvectors of the matrix $K_{1,kl}$ and chooses them as basis vectors then the
Neumann matrices are reduced to the diagonal form \cite{RSZ,BL,Erler,AG,MS}. The
calculations are greatly simplified in this basis \cite{RSZ,BL}.

 One
can search for eigenfunctions of $K_1$ on the complex plane $z$, but it is convenient
to use the map \cite{RSZ}
 \begin{equation}
  z=i\tanh w, \label{zw}
  \end{equation}
which takes the unit disk into the strip $|\text{Im}~w|\leq\pi/4$. In $w$ coordinate
the operator $K_1$ takes the form
\begin{equation}
 K_1=-i\frac{d}{dw}+i\tanh w.
\label{K_1}
\end{equation}
 The eigenvalues of the operator are all of the real
numbers $\kappa$. The eigenfunctions of \eqref{K_1} are
\begin{equation}
|\kappa,1/2\rangle(z)=\bigl[A_{1/2}(\kappa)\bigr]^{1/2}\, \cosh w(z)\,e^{i\kappa
w(z)}, \label{kappas}
\end{equation}
 where
$$A_{1/2}(\kappa)=\frac{1}{2\cosh\frac{\pi\kappa}{2}}$$
 is the normalization constant, it is determined from
\begin{equation}
\label{complit}
\langle\kappa,1/2|\kappa',1/2\rangle=\delta(\kappa-\kappa').
\end{equation}
The
transformation ``matrix'' between the discrete and continuous bases is
\begin{equation}
\label{<mk>} \langle
m,1/2|\kappa,1/2\rangle=V^{(1/2)}_m(\kappa)(A_{1/2}(\kappa))^{\frac12}.
\end{equation}
 Here the polynomials $V^{(1/2)}_m(\kappa)$ are given by the generating
function
\begin{equation}
 \cosh w\,e^{i\kappa w} =\sum_{m=0}^{\infty}
V_m^{(1/2)}(\kappa)z^m. \label{223}
\end{equation}
\subsection{Matter sector}
\subsubsection{Matter sector in the $\kappa$-basis}
Let $\psi(z)$ be a field of dimension $s=1/2$. In
the $NS$ sector it has a half-integer mode expansion. Following \cite{BL,AG,RZ} we
decompose it into creation and annihilation parts with respect to the
$SL(2,\mathbb{R})$-invariant vacuum $|0\rangle$:
\begin{eqnarray}
\psi(z)&=&\sum_{n=0}^{\infty}\left[\psi_{-(n+\frac12)}\,z^n+ \psi_{n+\frac12}
z^{-n-1} \right]\nonumber\\
&=&\sum_{n=0}^{\infty}\left[\psi_{-(n+\frac12)}\,|1/2,n\rangle(z)+
\psi_{n+\frac12}|1/2,n\rangle(1/z) z^{-1} \right]. \label{32}
\end{eqnarray} Using
the completeness condition
$$
\text{Id}(z,1/z)=\int_\mathbb{R}d\kappa|\kappa,1/2\rangle(z)\otimes\langle\kappa,1/2|(1/z)
$$
and (\ref{<mk>}), one gets
\begin{equation*}
\psi(z)=\int_\mathbb{R}d\kappa\sum_{n=0}^{\infty}(A_{1/2}(\kappa))^{\frac12}
\left[V^{(1/2)}_n(\kappa)\psi_{-(n+\frac12)}|\kappa,1/2\rangle(z)
+V^{(1/2)}_n(\kappa)\psi_{n+\frac12}|\kappa,1/2\rangle(1/z)z^{-1}\right],
\end{equation*}
or if one introduces the notation
\begin{equation}
\psi^{\pm}(\kappa)=\sqrt{A_{1/2}(\kappa)}\, \sum_{m=0}^{\infty} V_m^{(1/2)}(\kappa)\,
\psi_{\mp(m+1/2)}, \label{phipm}
\end{equation}
we have
\begin{equation*}
\psi(z)=\int_\mathbb{R}d\kappa \left[\psi^+(\kappa)|\kappa,1/2\rangle(z)
+\psi^-(\kappa)|\kappa,1/2\rangle(1/z)z^{-1}\right].
\end{equation*}
The
anticommutation relations between the oscillators in $\kappa$-basis are
\begin{equation}
 \{\psi^{-}(\kappa),\,\psi^+(\kappa')\}=\delta(\kappa-\kappa').
\end{equation}

\subsubsection{$\langle V_3^\psi|V_1^\psi\rangle$ in the $\kappa$-basis}

In this section we will consider the descent relation $\langle
V_3^\psi| V_1^\psi\rangle$ for the matter sector of the $NS$ string
in the $\kappa$-basis \cite{AG,MS}.

The right identity $| V_1^\psi\rangle$ in the $\kappa$-basis
\cite{AG} reads
\begin{equation}
\label{right_identity} |
V_1^\psi\rangle=\exp\left\{-\frac12\int_0^\infty
d\kappa\tau(\kappa)\psi_\alpha^+(\kappa)\varepsilon_{\alpha\beta}\psi_\beta^+(\kappa)\right\}|0\rangle
,\quad\alpha,\beta=1,2
 \end{equation}
  where $\tau(\kappa)=\tanh(\frac{\pi\kappa}{4})$
and $\psi_\alpha^+(\kappa)$ is the two component vector
\begin{eqnarray*}
\psi_1^\pm(\kappa)&=&\frac{1}{\sqrt2}(\psi^\pm(\kappa)-\psi^\pm(-\kappa)),\\
\psi_2^\pm(\kappa)&=&\frac{1}{\sqrt2}(\psi^\pm(\kappa)+\psi^\pm(-\kappa)),
\end{eqnarray*}
and $$ \varepsilon_{\alpha\beta}=\left( \begin{array}{cc}
  0 & 1 \\
  -1 & 0 \\
\end{array}\right). $$
The $|\,0\rangle$ stands for the usual oscillator vacuum, it doesn't
change under transformation from the discrete basis to the
$\kappa$-basis.

The three-string vertex is \cite{AG}
\begin{equation}
\label{three-string} \langle
V_3^\psi|={}_{123}\langle0|\exp\left\{-\frac12\int_0^\infty
d\kappa\psi_\alpha^{a-}(\kappa)V_{\alpha\beta}^{ab}(\kappa)\psi_\beta^{b-}(\kappa)\right\},\quad
a,b=1,2,3
\end{equation}
where $V_{\alpha\beta}^{ab}(\kappa)$ is $6\times6$ matrix
defined by
\begin{equation}
\label{V}
V_{\alpha\beta}^{ab}(\kappa)=\mu(\kappa)\varepsilon_{\alpha\beta}\otimes\delta^{ab}+
\mu_t(\kappa)c_{\alpha\beta}\otimes\chi^{ab}+\mu_s(\kappa)\varepsilon_{\alpha\beta}\otimes\epsilon^{ab};
\end{equation}
\begin{equation*}
\mu=\tau\frac{1-\tau^2}{1+3\tau^2},\quad\mu_t=\frac{1+\tau^2}{1+3\tau^2},\quad\mu_s=
-\tau\frac{1+\tau^2}{1+3\tau^2},
\end{equation*}
 and $$ c_{\alpha\beta}=\left(%
\begin{array}{cc}
  1 & 0 \\
  0 & -1 \\
\end{array} \right). $$

We use the following notations \cite{BK}
$$ \chi^{ab}=\left(\begin{array}{ccc}
  0 & 1 & -1 \\
  -1 & 0 & 1 \\
  1 & -1 & 0 \\
\end{array} \right),\qquad\epsilon^{ab}=\left( \begin{array}{ccc}
  0 & 1 & 1 \\
  1 & 0 & 1 \\
  1 & 1 & 0 \\
\end{array} \right).
$$ Hence, all the vertices  are defined. We can evaluate the
descent relation
\begin{eqnarray*}
\langle V_3^\psi|V_1^\psi\rangle&=&\\
&=&{}_{123}\langle0|\exp\left\{-\frac12\int_0^\infty
d\kappa\psi_\alpha^{a-}(\kappa)V_{\alpha\beta}^{ab}(\kappa)\psi_\beta^{b-}(\kappa)\right\}\\
&\times&\exp\left\{-\frac12\int_0^\infty
d\kappa'\tau(\kappa')\psi_\alpha^{1+}(\kappa')\varepsilon_{\alpha\beta}\psi_\beta^{1+}(\kappa')\right\}
|0\rangle_1,
 \end{eqnarray*}
 we use the formula of multiplication of two exponents in
\cite{AG,AGM}
\begin{eqnarray*}
 &=&\det(1-\mu\tau(\kappa))^{10}\\
&\times&{}_{23}\langle0|\exp\left\{-\frac12\int_0^\infty
d\kappa\psi_\alpha^{n-}(\kappa)
\left[V_{\alpha\beta}^{nm}(\kappa)-V_{\alpha\gamma}^{n1}(\kappa)\frac{\tau(\kappa)}{1-\mu\tau(\kappa)}
\varepsilon_{\gamma\delta}V_{\delta\beta}^{1m}(\kappa)\right]
\psi_{\beta}^{m-}(\kappa)\right\}.
 \end{eqnarray*}
 Here
$\det(1-\mu\tau(\kappa))^{10}$ stands for the determinated of the diagonal operator
$(1-\mu\tau(\kappa))\delta(k-k')$.

Let us introduce the notation
\begin{equation}
U_{\alpha\beta}^{nm}(\kappa)=
V_{\alpha\beta}^{nm}(\kappa)-V_{\alpha\gamma}^{n1}(\kappa)\frac{\tau(\kappa)}{1-\mu\tau(\kappa)}
\varepsilon_{\gamma\delta}V_{\delta\beta}^{1m}(\kappa),\qquad m=2,3.
\end{equation}
It is easy to check using the explicit expression of the matrix $V^{ab}(\kappa)$
(\ref{V}) that the matrix $U_{\alpha\beta}^{nm}(\kappa)$ has the form
\begin{equation} \label{conditions}
U_{\alpha\beta}^{22}(\kappa)=U_{\alpha\beta}^{33}(\kappa)=0,\quad\text{and}\quad
U_{\alpha\beta}^{23}(\kappa)-U_{\beta\alpha}^{32}(\kappa)=2c_{\alpha\beta}.
\end{equation}
Thus we get the following form for the vertex $\langle V_2|$
\begin{equation}\label{two-string} \langle
V_2^\psi|={}_{12}\langle0|\exp\left\{-\int_0^\infty
d\kappa\psi_\alpha^{1-}(\kappa)c_{\alpha\beta}\psi_\beta^{2-}(\kappa)\right\}.
\end{equation}

Thereby,
\begin{eqnarray} \langle V_3^\psi|
V_1^\psi\rangle&=&\det(1-\mu\tau(\kappa))^{10}{}_{23}\langle0|\exp\left\{-\int_0^\infty
d\kappa\psi_\alpha^{2-}(\kappa)c_{\alpha\beta}
\psi_\beta^{3-}(\kappa)\right\}\nonumber\\
&=&\det(1-\mu\tau(\kappa))^{10}\langle V_2^\psi|.
 \end{eqnarray}
 We
got the vertex $\langle V_2^\psi|$ and
$\det(1-\mu\tau(\kappa))^{10}$ which is obviously divergent. Below
we will show how to treat the determinant following \cite{BL}.

To calculate $\det A$ one uses the standard trick
\begin{equation*}
\det A=\exp
\{\text{Tr}\log A\}.
\end{equation*}
Let us suppose that we have some operator $A$
which is diagonal in the $\kappa$-basis and has eigenvalues $A(\kappa)$
\begin{equation}
\text{Tr}~A=\int_0^\infty
d\kappa\langle\kappa,1/2|A|1/2,\kappa\rangle=\int_0^\infty d\kappa
A(\kappa)\langle\kappa,1/2|1/2,\kappa\rangle.
\end{equation}
 Taking into account
(\ref{complit}) one gets
\begin{equation}
\label{A} \text{Tr}A=\int_0^\infty d\kappa
A(\kappa)\langle\kappa,1/2|1/2,\kappa\rangle\sim\delta(0)\int_0^\infty d\kappa~
A(\kappa).
 \end{equation}
  This expression diverges. One would like to regularize
$\delta(0)$ and at first sight for this one can use any regularization of delta
function. In \cite{Belov2,BL} the arguments were given how to choose the
regularization. It was suggested to regularize the measure in the inner product
(\ref{norm}) by $s\to s+\Delta$. The main argument for the regularization of measure
was given through the associativity of operators $U_p$ which add the nonzero momentum
to vertices. In \cite{Manes} it was suggested that the anomaly in the associativity
$$ \langle V_N|U_{p+q}=\langle V_N|(U_pU_q)\neq(\langle V_N|U_p)U_q $$
is connected
with the breaking of the unitarity of the operator $U_p$. In order to act by the
operator $U_p$ on a vertex it should be regularized. The regularization could break
the unitarity. Manes used the level truncation method. Next in \cite{Potting} the
same calculation was done, but they used the regularization with the
$\zeta$-function. They concluded that there is a regularization free from the
anomaly. They suggested the complicated regularization in which the anomaly was
absent. Using the $\Delta$ regularization of \cite{Belov2} it is easy to prove the
associativity
\begin{equation}
\lim_{\Delta\to+0}\langle
V_N|U_{q+p}^{(\Delta)}=\lim_{\Delta\to+0}\left[\lim_{\Delta'\to+0}\langle
V_N|U_q^{(\Delta')}\right]U_p^{(\Delta)}.
 \end{equation}
 It was suggested
to introduce the spectral regularized density as
 \begin{equation}
\rho_{s,\Delta}(\kappa',\kappa)\equiv\langle\kappa',s|\kappa,s\rangle_{s+\Delta}
\end{equation}
and
\begin{equation}
\lim_{\Delta\to0}\rho_{s,\Delta}(\kappa',\kappa)=\rho_s(\kappa',\kappa)=\delta(\kappa-\kappa').
\end{equation}
 Thus the regularized expression (\ref{A}) can be written as
\begin{equation}
\text{Tr}_\Delta~A=\int_0^\infty d\kappa
A(\kappa)\rho_{1/2,\Delta}(\kappa,\kappa).
 \end{equation}
 So one can calculate the
determinant of the operator through the trace of the operator with a regularized
spectral density. The expression of the regularized spectral density for an arbitrary
$s$ was found in \cite{Belov2}
\begin{eqnarray}\label{ro}
\rho_{s,\Delta}(\kappa)&=&\frac{1}{4\pi\Delta}+\frac{\log2}{2\pi}
-B_s(\kappa),\nonumber\\
\end{eqnarray}
 where
 \begin{equation*}
B_s(\kappa)=\frac{1}{4\pi}
\left[\psi\left(s+\frac{i\kappa}{2}\right)+\psi\left(s-\frac{i\kappa}{2}\right)\right],
\end{equation*}
and $\psi(z)$ is the logarithmic derivation of the $\Gamma$-function.
The spectral density was also calculated for $s=0$ in \cite{FKM}.

Taking into account the discussion given above we can calculate the determinant
\begin{equation*}
\det(1-\mu\tau(\kappa))^{10}=e^{10\text{Tr}\log(1-\mu\tau(\kappa))},
\end{equation*}
where
\begin{equation*}
\text{Tr}\log(1-\mu\tau(\kappa))=\lim_{\Delta\to0}\text{Tr}_\Delta\log(1-\mu\tau(\kappa))
\end{equation*}
and
\begin{equation}\label{Tr}
\text{Tr}_\Delta\log(1-\mu\tau(\kappa))=\int_0^\infty
d\kappa\log(1-\mu\tau(\kappa))\rho_{1/2,\Delta}(\kappa).
\end{equation}
\subsection{Superghost sector}
\subsubsection{Superghost sector in the $\kappa$-basis}

Since the Neumann matrices of vertices for superghosts are the
functions of the Neumann matrices of vertices for the matter, they
have  the general system of the eigenvectors. Therefore one can
introduce the superghost oscillators in the $\kappa$-basis in the
following way \cite{AG}: \begin{equation}
\beta^{\pm}(\kappa)=\pm\sqrt{A_{1/2}(\kappa)}\, \sum_{m=0}^{\infty}
V_m^{(1/2)}(\kappa)\, \beta_{\mp(n+1/2)}
 \end{equation}
  and
\begin{equation}
\gamma^{\pm}(\kappa)=\sqrt{A_{1/2}(\kappa)}\, \sum_{m=0}^{\infty}
V_m^{(1/2)}(\kappa)\, \gamma_{\mp(n+1/2)}.
 \end{equation}
 As a consequence of the commutation relation in the discrete basis
 $[\gamma_r,\beta_s]=\delta_{r+s,0}$, we have the following commutation relation
 in the continuous basis
\begin{equation*} [\gamma^-(\kappa),\beta^+(\kappa')]=\delta(\kappa-\kappa').
\end{equation*}
\subsubsection{$\langle V_3^{\beta\gamma}|V_1^{\beta\gamma}\rangle$ in the $\kappa$-basis}
Now we are able to calculate the descent relation for the
superghosts in the $\kappa$-basis.

The vertices in the $\kappa$-basis have the form \cite{AG}
\begin{equation}
|V_1^{\beta\gamma}\rangle=e^{\phi(\frac{\pi}{2})}\exp\left\{\int_0^\infty
d\kappa\widetilde{\tau}(\kappa)\beta_\alpha^+(\kappa)
\varepsilon_{\alpha\beta}\gamma_\beta^+(\kappa)\right\}|-1\rangle
\end{equation} and
\begin{equation} \langle
V_3^{\beta\gamma}|={}_{123}\langle-1|\exp\left\{-\int_0^\infty
d\kappa\beta_\alpha^{a-}(\kappa)
K_{\alpha\beta}^{ab}(\kappa)\gamma_\beta^{b-}(\kappa)\right\}e^{-\phi^1(\frac\pi2)},
\end{equation}
 where $\alpha,\beta$~= 1,2 and $a,b$~= 1,2,3. As in the case of the
fermionic matter the superghosts $\beta_\alpha^\pm(\kappa)$ and
$\gamma_\beta^\pm(\kappa)$ are two component vectors:
\begin{eqnarray*}
\beta_1^\pm(\kappa)=\frac{1}{\sqrt2}(\beta^\pm(\kappa)-\beta^\pm(-\kappa)), &\quad&
\beta_2^\pm(\kappa)=\frac{1}{\sqrt2}(\beta^\pm(\kappa)+\beta^\pm(-\kappa)),\\
\gamma_1^\pm(\kappa)=\frac{1}{\sqrt2}(\gamma^\pm(\kappa)-\gamma^\pm(-\kappa)),
&\quad&
\gamma_2^\pm(\kappa)=\frac{1}{\sqrt2}(\gamma^\pm(\kappa)+\gamma^\pm(-\kappa)).
\end{eqnarray*}
 The matrix $K_{\alpha\beta}^{ab}(\kappa)$ is
\begin{equation}\label{K}
K_{\alpha\beta}^{ab}(\kappa)=\widetilde{\mu}(\kappa)\varepsilon_{\alpha\beta}\otimes\delta^{ab}+
\widetilde{\mu}_t(\kappa)c_{\alpha\beta}\otimes\chi^{ab}+\widetilde{\mu}_s(\kappa)
\varepsilon_{\alpha\beta}\otimes\epsilon^{ab}; \end{equation} where
\begin{equation*} \widetilde{\mu}=
\widetilde{\tau}\frac{\widetilde{\tau}^2-1}{1+3\widetilde{\tau}^2}\quad
\widetilde{\mu}_t=-\frac{1+\widetilde{\tau}^2}{1+3\widetilde{\tau}^2},
\quad\widetilde{\mu}_s=
\widetilde{\tau}\frac{1+\widetilde{\tau}^2}{1+3\widetilde{\tau}^2},
\end{equation*} the function
$\widetilde{\tau}(\kappa)=\coth(\frac{\pi\kappa}{4})$.

Hence, the descent relation in the $\kappa$-basis is
\begin{eqnarray} \label{descent-rel-kappa} \langle
V_3^{\beta\gamma}|
V_1^{\beta\gamma}\rangle&=&{}_{123}\langle-1|\exp\left\{-\int_0^\infty
d\kappa\beta_\alpha^{a-}(\kappa)
K_{\alpha\beta}^{ab}(\kappa)\gamma_\beta^{b-}(\kappa)\right\}e^{-\phi^1(\frac\pi2)}\nonumber\\
&\times&e^{\phi^1(\frac\pi2)}\exp\left\{\int_0^\infty
d\kappa'\widetilde{\tau}(\kappa')\beta_\alpha^{1+}(\kappa')
\varepsilon_{\alpha\beta}\gamma_\beta^{1+}(\kappa')\right\}|-1\rangle_1\\
 &=&{}_{123}\langle-1|\exp\left\{-\int_0^\infty
d\kappa\beta_\alpha^{a-}(\kappa)
K_{\alpha\beta}^{ab}(\kappa)\gamma_\beta^{b-}(\kappa)\right\}\nonumber\\
&\times&\exp\left\{\int_0^\infty
d\kappa'\widetilde{\tau}(\kappa')\beta_\alpha^{1+}(\kappa')
\varepsilon_{\alpha\beta}\gamma_\beta^{1+}(\kappa')\right\}|-1\rangle_1,
\end{eqnarray}

 Taking into account the expression for the inner product of two
exponents \cite{Belov2,AGK} the descent relation
(\ref{descent-rel-kappa}) has the form ($n,m=2,3$) \begin{equation}
=\det(1+\widetilde{\tau}\widetilde{\mu}(\kappa))^{-2}{}_{23}\langle-1|
\exp\left\{-\int_0^\infty d\kappa\beta_\alpha^{n-}(\kappa)
\widetilde{U}_{\alpha\beta}^{nm}(\kappa)\gamma_\beta^{m-}(\kappa)\right\},
\end{equation}
where
\begin{equation} \widetilde{U}_{\alpha\beta}^{nm}(\kappa)\equiv
K_{\alpha\beta}^{nm} (\kappa)+K_{\alpha\gamma}^{n1}(\kappa)
\frac{\widetilde{\tau}(\kappa)}{1+\widetilde{\mu}\widetilde{\tau}(\kappa)}
\varepsilon_{\gamma\delta}K_{\delta\beta}^{1m}(\kappa).
\end{equation}
It is easy to
check that the matrix $\widetilde{U}_{\kappa,\alpha\beta}^{nm}$ satisfies the
following conditions
\begin{equation} \label{conditions2}
\widetilde{U}_{\alpha\beta}^{22}(\kappa)=\widetilde{U}_{\alpha\beta}^{33}(\kappa)=0\quad\text{and}\quad
\widetilde{U}_{\alpha\beta}^{23}(\kappa)=-c_{\alpha\beta},\quad
\widetilde{U}_{\beta\alpha}^{32}(\kappa)=c_{\alpha\beta}.
 \end{equation}
  Hence, the
two-string vertex in the $\kappa$-basis has the form
 \begin{equation}
  \langle
V_2^{\beta\gamma}|={}_{23}\langle-1|\exp\left\{\int_0^\infty
d\kappa\left(\beta_\alpha^{2-}(\kappa)c_{\alpha\beta}\gamma_\beta^{3-}(\kappa)-
\beta_\alpha^{3-}(\kappa)c_{\alpha\beta}\gamma_\beta^{2-}(\kappa)\right)\right\}.
\end{equation}
However, we have expected another form for the vertex $\langle
V_2^{\beta\gamma}|$. We have thought that the vertex $\langle V_2|$
in the $\kappa$-basis looks like (\ref{V2}).

Below we give comments on this result together with the result of calculation in the
discrete basis.

Following the discussion above we get the
 following result
\begin{equation} \langle V_3^{\beta\gamma}| V_1^{\beta\gamma}\rangle=
\det(1+\widetilde{\tau}\widetilde{\mu}(\kappa))^{-2}\langle V_2^{\beta\gamma}|.
\end{equation} As above we represent
$\det(1+\widetilde{\tau}\widetilde{\mu}(\kappa))$ as \begin{equation*}
\det(1+\widetilde{\mu}\widetilde{\tau}(\kappa))=e^{Tr\log(1+\widetilde{\mu}\widetilde{\tau}(\kappa))},
\end{equation*} and as above we regularize it by regularization of the measure
\begin{equation} Tr_\Delta\log(1+\widetilde{\mu}\widetilde{\tau}(\kappa))
=\int_0^\infty
d\kappa\log(1+\widetilde{\mu}\widetilde{\tau}(\kappa))\rho_{1/2,\Delta}(\kappa).
\end{equation}
\subsection{Normalization factor $\mathcal{Z}_3$}

The descent relation in the $NS$ sector of superstring is (the descent relation in
the bosonic string in the $\kappa$-basis was calculated in \cite{Belov2})
\begin{equation}
 \langle V_3|V_1\rangle=\mathcal{Z}_3\langle
V_2|=Z^xZ^{bc}Z^\psi Z^{\beta\gamma}\langle V_2|,
 \end{equation}
 where
\begin{equation}\label{123}
\log\mathcal{Z}_3=-11\log\det(1-\nu)+F_{3,1}+F_{1,1}-F_{2,1}+10\log\det(1-\mu\tau)
-2\log\det(1+\widetilde{\mu}\widetilde{\tau}) \end{equation} and
$F_{N,1}$ was defined in \cite{Belov2}. \begin{equation}
F_{N,1}\equiv\frac92\frac{(N-2)^2}{2N}\left(\frac{1}{4\Delta}+\frac{\gamma_E-\log2}{2}\right)+
\frac92\left[\log\frac{N}{2}-\frac{N-2}{2}\log2\right],\qquad
N=1,2,3. \end{equation} The function $\nu(\kappa)$ is \cite{BL}
\begin{equation*}
\nu(\kappa)=-\frac{\sinh(\frac{\pi\kappa}{4})}{\sinh(\frac{3\pi\kappa}{4})}.
\end{equation*}
 In other words
$\log\mathcal{Z}_3$ is \begin{eqnarray}\label{sing}
\log\mathcal{Z}_3&=&-11\int_0^\infty
d\kappa\log(1-\nu)\rho_{1,\Delta}(\kappa)+3\left(\frac{1}{4\Delta}+\frac{\gamma_E-\log2}{2}\right)
+\frac92\log\frac34\nonumber\\ &&+10\int_0^\infty
d\kappa\log(1-\mu\tau)\rho_{1/2,\Delta}(\kappa)-2\int_0^\infty
d\kappa\log(1+\widetilde{\mu}\widetilde{\tau})\rho_{1/2,\Delta}(\kappa),
\end{eqnarray} here $\gamma_E$ is the Euler constant.

Now, we extract the $1/\Delta$ terms from (\ref{sing})
\begin{equation}\label{singul} \left(3-\frac{11}{\pi}\int_0^\infty
d\kappa\log(1-\nu)+\frac{10}{\pi}\int_0^\infty
d\kappa\log(1-\mu\tau)-\frac2\pi\int_0^\infty
d\kappa\log(1+\widetilde{\mu}\widetilde{\tau})\right)\frac{1}{4\Delta}.
\end{equation} The integrals in (\ref{singul}) are easy to calculate:
\begin{equation*} \int_0^\infty d\kappa\log(1-\nu)=\frac\pi9,\quad \int_0^\infty
d\kappa\log(1-\mu\tau)=-\frac{\pi}{18},\quad \int_0^\infty
d\kappa\log(1+\widetilde{\mu}\widetilde{\tau})=\frac{11\pi}{18}. \end{equation*} The
simple algebra gives the cancelation of the singular part (\ref{sing})
\begin{equation}
\text{singular
part}=\left(3-\frac{11}{9}-\frac{10}{18}-\frac{22}{18}\right)\frac{1}{4\Delta}
=\left(3-\frac{54}{18}\right)\frac{1}{4\Delta}=0.
 \end{equation} Thereby, the
$1/\Delta$ part of (\ref{sing}) is zero in critical dimension that
enter explicitly in eq.(\ref{123}).

The rest in (\ref{sing}) is (see (\ref{ro}))
\begin{eqnarray}\label{finite}
\text{finite\,part}&=&\frac32\log\frac{27}{256}+\frac{3\gamma_E}{2}+
11\int_0^\infty d\kappa\log(1-\nu)B_1(\kappa)\nonumber\\ &-&
10\int_0^\infty d\kappa\log(1-\mu\tau)B_{1/2}(\kappa)+
2\int_0^\infty
d\kappa\log(1+\widetilde{\mu}\widetilde{\tau})B_{1/2}(\kappa)
\end{eqnarray}
These integrals can be calculate analytically following the lines of
\cite{BL,FKM} \footnote{We are grateful to our referee for the
careful explanation of the calculation procedure.}. Due to the
analytic calculations we get the following value for the factor
$\mathcal{Z}_3$
\begin{equation}
\mathcal{Z}_3\approx0,02.
\end{equation}
 Below we give comments on this result together with
the result of the calculation in the discrete basis.
\section{ Descent relation in
the discrete basis}
In this section we evaluate the descent relation $\langle V_3|
V_1\rangle$ for the $NS$ string fermionic in the matter and ghost
sectors in the discrete basis.
\subsection{Matter sector}

The three-string vertex $\langle V_3^\psi|$ is \cite{GJ} \begin{equation} \langle
V_3^\psi|={}_{321}\langle0|\exp\left\{-\frac12\sum_{r,s\geq1/2}^\infty\psi_r^
aV^{ab}_{rs}\psi_s^b\right\}, \end{equation} where $a,b=1,2,3$.

The Neumann matrices $V^{ab}$ were built in \cite{GJ} with using the Neumann function
method. The more convenient representation for these matrices was developed at
\cite{MS,AGM}.

The LHS of the descent relation (we drop the index $\mu$) reads
\begin{equation}
\label{descent_relation} \langle V_3^\psi| V_1^\psi\rangle=
{}_{321}\langle0|e^{-\frac12\psi_r^aV^{ab}_{rs}\psi_s^b}
e^{\frac12\psi_{-r}^1 I_{rs}\psi_{-s}^1}|0\rangle_1.
\end{equation}
Let us mark out the index ``1''. For this we rewrite the expression
$\psi_r^aV^{ab}_{rs}\psi_s^b$ as \begin{equation*}
\psi_r^1V^{11}_{rs}\psi_s^1+2\psi_r^1V^{1q}_{rs}\psi_s^q+
\psi_r^pV^{pq}_{rs}\psi_s^q,\quad p,q=2,3, \end{equation*} here we
use the following properties of Neumann matrices \begin{equation*}
V^{ab}_{rs}=-V^{ba}_{sr}.
\end{equation*} It is useful to introduce the following notation $$ \lambda_r\equiv
V^{1q}_{rs}\psi_s^q. $$ Thereby, (\ref{descent_relation})  can be rewritten as
$(p,q=2,3)$
\begin{equation} \langle
V_3^{\psi}|V_1^{\psi}\rangle={}_{32}\langle0|e^{-\frac12\psi_r^pV^{pq}_{rs}\psi_s^q}{}_1
\langle0|e^{-\frac12\psi_r^1V^{11}_{rs}\psi_s^1-\psi_r^1\lambda_r}
e^{\frac12\psi_{-r}^1 I_{rs}\psi_{-s}^1}|0\rangle_1.
\end{equation}
 Next using the
identity (\ref{ff}) the descent relation (\ref{descent_relation}) can be written as
\begin{equation} \langle V_3^\psi|
V_1^\psi\rangle=\det(1-V^{11}I)^{10/2}{}_{32}\langle0|\exp
\left\{-\frac12\psi_r^pR^{pq}_{rs}\psi_s^q\right\},
\end{equation}
 where
\begin{equation*} R^{pq}_{rs}\equiv
V^{pq}_{rs}+V^{p1}_{rk}((1-IV^{11})^{-1}I)_{kl}V^{1q}_{ls}.
\end{equation*}
We know
that the two-string vertex is \cite{GJ}
 \begin{equation} \langle
V_2^\psi|={}_{23}\langle0|\exp\{-i\sum_{r\geq1/2}\psi^2_r(-)^r\psi^3_r\}.
\end{equation}
 Therefore, the matrices $R^{pq}_{rs}$ should have the following form
to satisfy the descent relation
\begin{equation}\label{con1} R^{pp}_{rs}=0,\quad
R^{23}_{rs}=i(-)^r\delta_{r,s},\quad R^{32}_{rs}=-i(-)^r\delta_{r,s}.
 \end{equation}

 We prove these conditions analytically (the details of this
calculation are presented in Appendix A). The fact that we got these results
analytically is very unexpected and remarkable, because the calculations of the
descent relation in the bosonic string demanded the numerical calculations in order
to provide the correct structure of the exponent in the vertex $\langle V_2|$. There
was no chance to make the calculations analytically because of the complicated
structure of the Neumann matrices in the vertices of the bosonic string.
\subsection{Superghost sector}
The vertex $| V_1^{\beta\gamma}\rangle$ can be presented in the form
(here we use another form for the vertex $|V_1^{\beta\gamma}\rangle$
which also was suggested in \cite{GJ}, moreover exactly this vertex
was written in the $\kappa$-basis)
\begin{equation}
|V_1^{\beta\gamma}\rangle=e^{\phi(\frac{\pi}{2})}
\exp\left\{\sum_{r,s\geq1/2}\beta_{-r}\tI_{rs}\gamma_{-s}\right\}|-1\rangle.
\end{equation}
Note that, we use another form for the vertex $|
V_1^{\beta\gamma}\rangle$ here. Exactly this form will be useful for
given calculation.

The three-string vertex is \cite{GJ}
\begin{equation} \langle
V_3^{\beta\gamma}|={}_{123}\langle-1|\exp\left\{-\sum_{a,b=1}^3\sum_{r\geq1/2\atop
s\geq1/2}\beta_r^a
K^{ab}_{rs}\gamma_s^b\right\}e^{-\phi^1(\frac\pi2)}.
\end{equation}

The ghost part of the LHS in the descent relation reads
\begin{eqnarray}
\label{spusc_new} &\langle V_3^{\beta\gamma}|
V_1^{\beta\gamma}\rangle&=\\
&=&{}_{123}\langle-1|\exp\left\{-\sum_{r\geq1/2\atop
s\geq1/2}\beta_r^a
K^{ab}_{rs}\gamma_s^b\right\}e^{-\phi(\frac\pi2)}e^{\phi(\frac\pi2)}\exp\left\{\sum_{r\geq1/2\atop
s\geq1/2}\beta_{-r}^1\tI_{rs}\gamma_{-s}^1\right\}|-1\rangle_1\nonumber
\end{eqnarray}
or
\begin{equation}\label{spusc_new_2}
{}_{123}\langle-1|\exp\left\{-\sum_{r\geq1/2\atop s\geq1/2}\beta_r^a
K^{ab}_{rs}\gamma_s^b\right\}\exp\left\{\sum_{r\geq1/2\atop
s\geq1/2}\beta_{-r}^1\tI_{rs}\gamma_{-s}^1\right\}|-1\rangle_1.
\end{equation}
It is
easy to evaluate (\ref{spusc_new_2}) using the identity (\ref{gg}). So we have the
following expression
\begin{equation}
 \langle V_3^{\beta\gamma}|
V_1^{\beta\gamma}\rangle=\det(1- K^{11}
\tI)^{-1}{}_{23}\langle-1|\exp\left\{-\sum_{r,s\geq1/2}\beta_r^pU_{rs}^{pq}\gamma_s^q\right\},\quad
p,q=2,3
 \end{equation}
  where
  \begin{equation} U_{rs}^{pq}\equiv K_{rs}^{pq}+
K_{rk}^{p1}((1- \tI K^{11})^{-1} \tI)_{kl}K_{ls}^{1q}.
\end{equation}
 In Appendix B
we prove analytically that the matrix $U_{rs}^{ab}$ satisfies the conditions
\begin{equation}\label{con2}
U_{rs}^{23}=-i(-)^r\delta_{r,s},\,\,U_{rs}^{32}=i(-)^r\delta_{r,s},\,\,U_{rs}^{22}=
U_{rs}^{33}=0,\,\,\text{for}\,\, r,s\geq1/2. \end{equation} Hence the two-string
vertex has the form
\begin{equation}
 \langle V_2^{\beta\gamma}|={}_{12}\langle-1|
\exp\left\{i\sum_{r\geq1/2}\beta_r^1(-)^r\gamma_r^2-i\sum_{r\geq1/2}\beta_r^2(-)^r\gamma_r^1\right\}.
\end{equation}
Above we calculated the descent relation in the discrete basis. We
used the fact that the vertex is factorized into the vertices of the
bosonic and fermionic matter and their ghosts and we checked the
descent relations separately for each vertex and got four
coefficients $Z$. In the total descent relation we got the product
of all these factors $\mathcal{Z}^{osc}_3=Z^X Z^{bc} Z^{\psi}
Z^{\beta\gamma}$. The bosonic \cite{KF,AGMR} and fermionic parts of
this coefficient read
\begin{eqnarray}\label{Z}
Z^x Z^{bc}&=&\frac i2\det(1+V^{11m}S)^{-5}\det(1-SX)J_0N_3,\nonumber\\
Z^\psi Z^{\beta\gamma}&=&\det(1-V^{11}I)^5\det(1-K^{11} \tI)^{-1}.
\end{eqnarray}

\begin{figure}[h!]
\begin{center}
\includegraphics[scale=0.45]{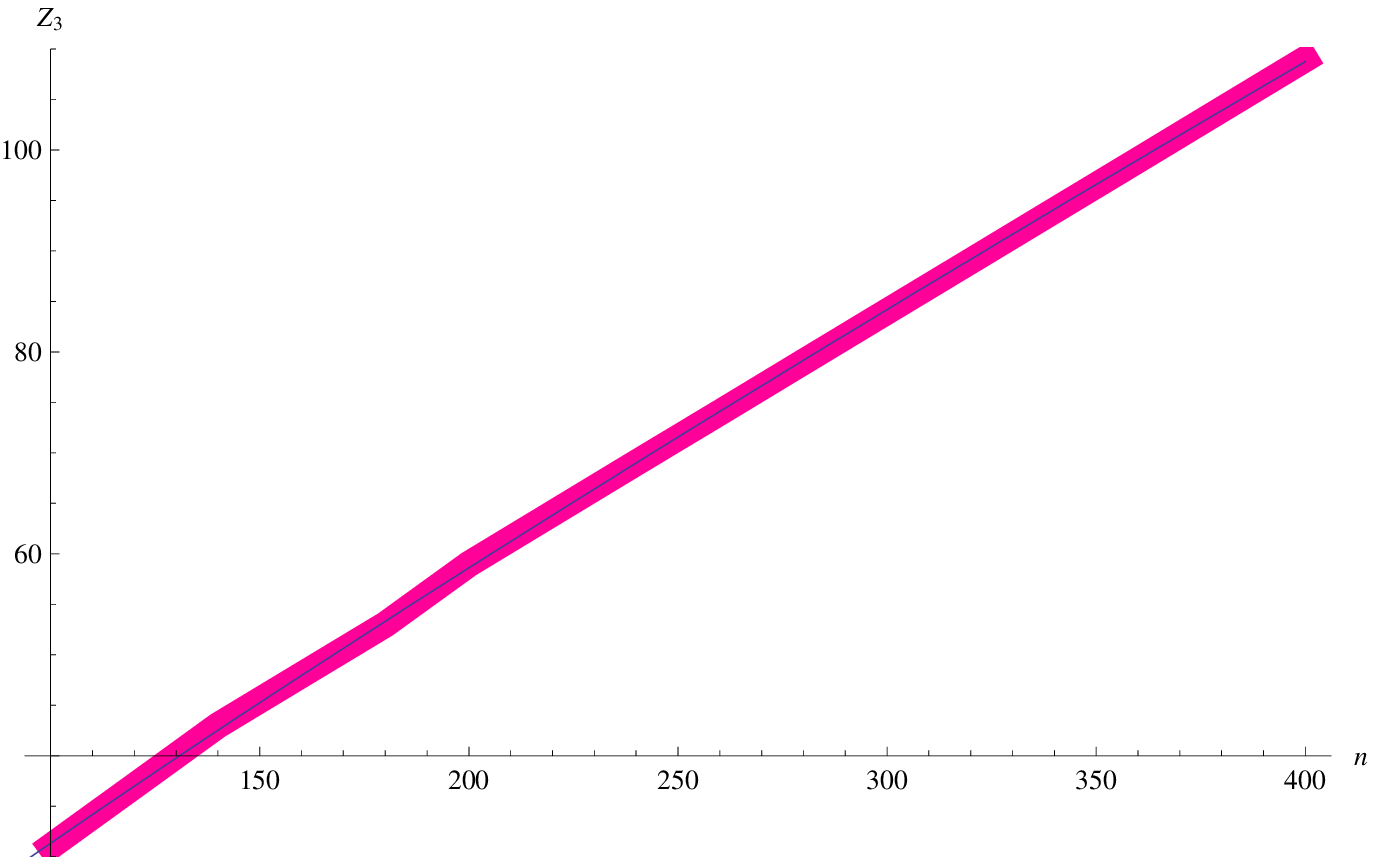}$\,\,\,$a)
\qquad
\includegraphics[scale=0.65]{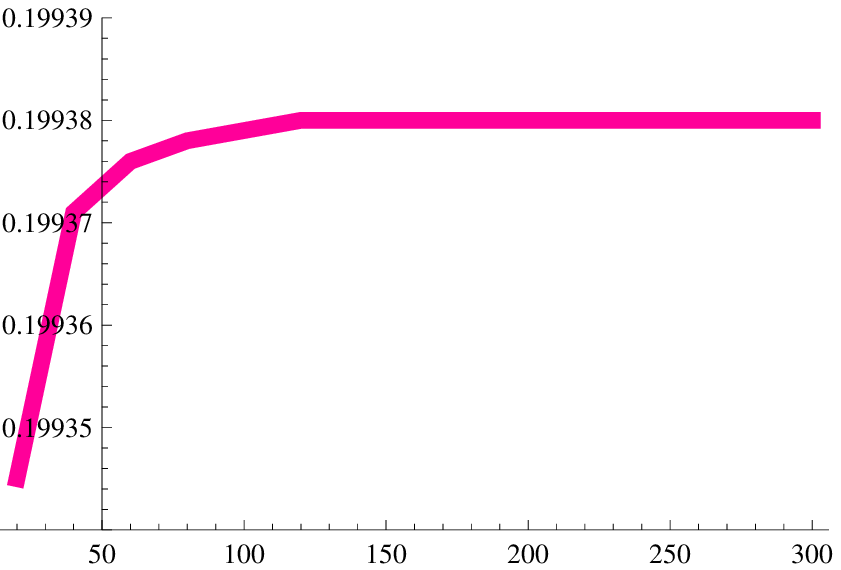}$\,\,\,$b)
\caption{a.\,$Z^{X}_3Z^{b,c}_3$ for D=10,$\,\,\,$
b.\,$Z^{X}_3Z^{b,c}_3$ for D=26 } \label{graph1}
\end{center}
\end{figure}

\begin{figure}[h!]
\begin{center}
\includegraphics[scale=0.6]{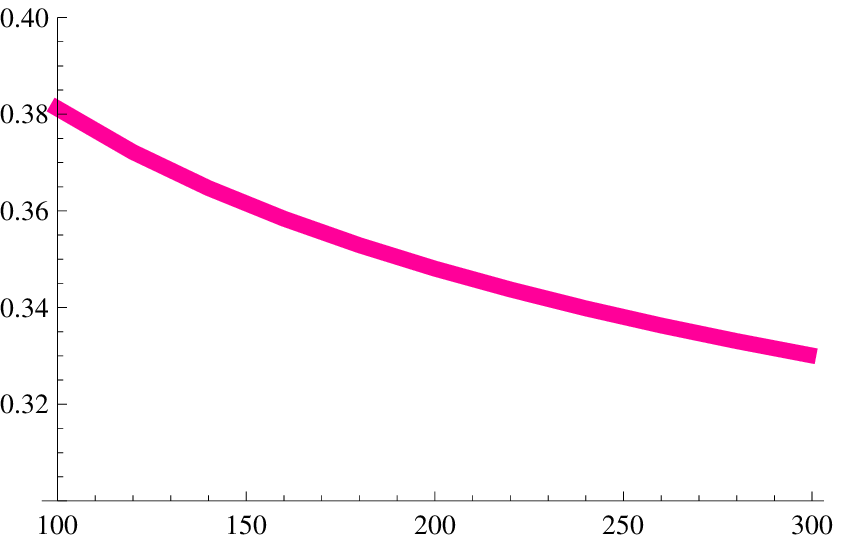}$\,\,\,$a)
\qquad\qquad
\includegraphics[scale=0.6]{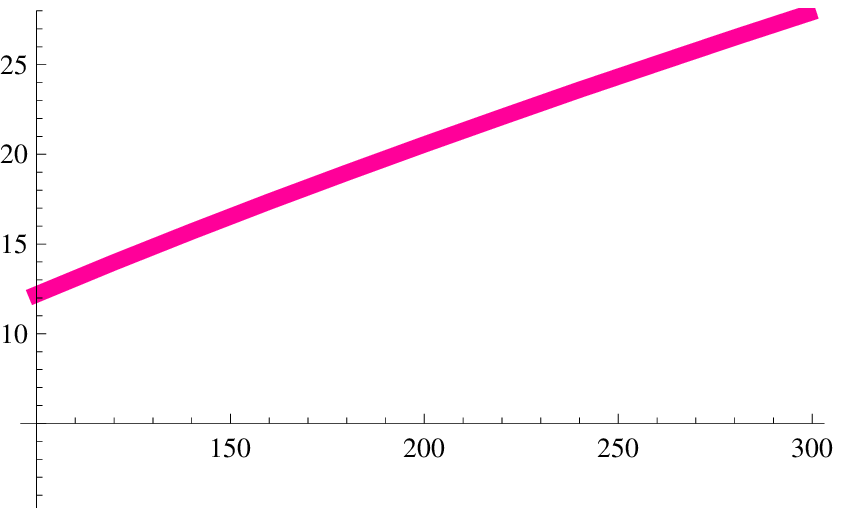}$\,\,\,$f)\\
\includegraphics[scale=0.4]{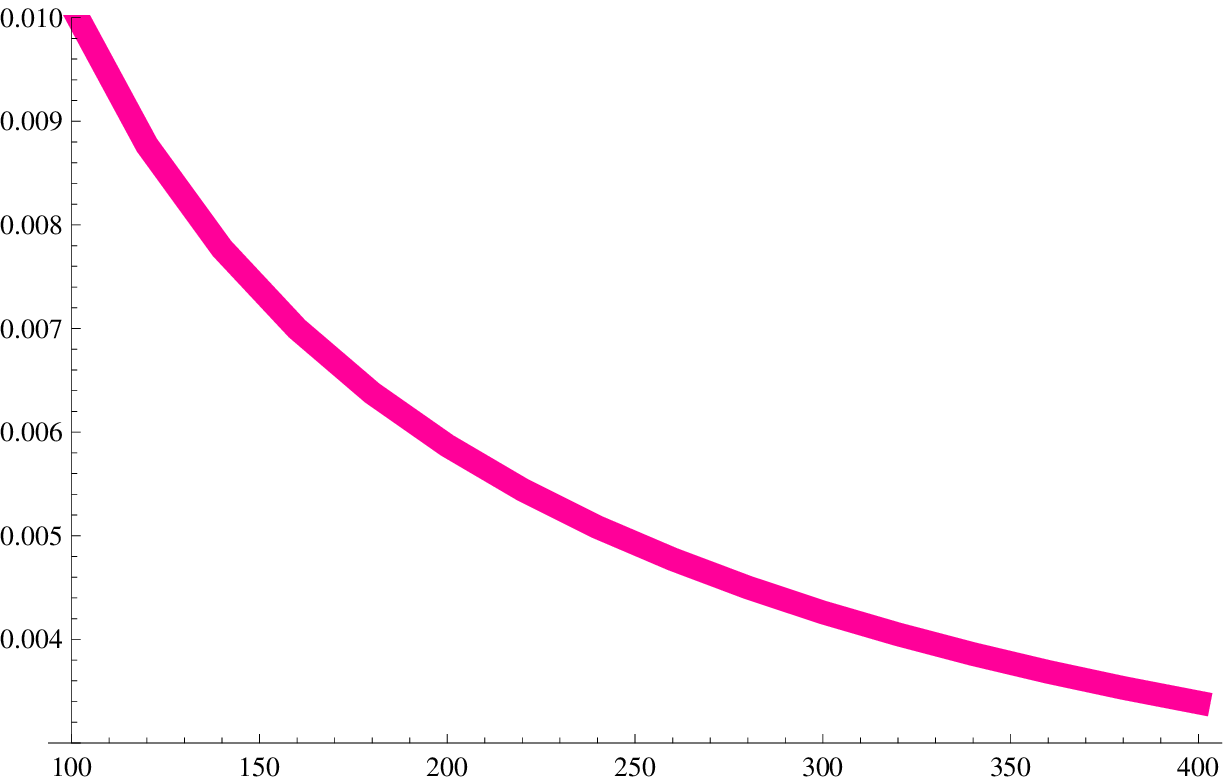}$\,\,\,$b)
\qquad\qquad
\includegraphics[scale=0.6]{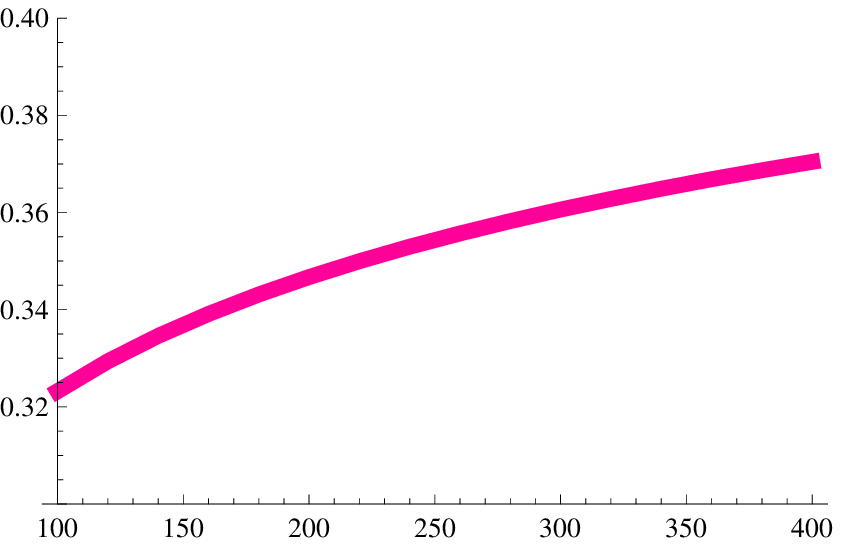}$\,\,\,$g)\\
\includegraphics[scale=0.6]{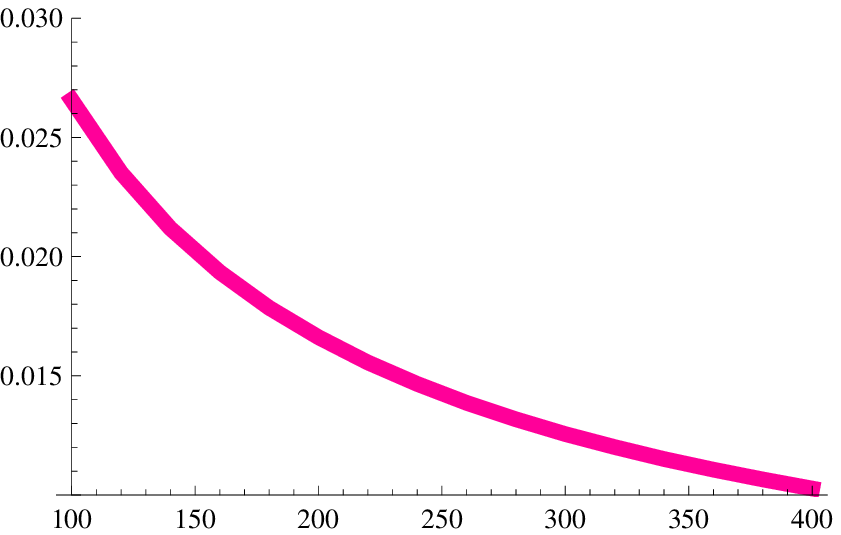}$\,\,\,$c)
\qquad\qquad
\includegraphics[scale=0.6]{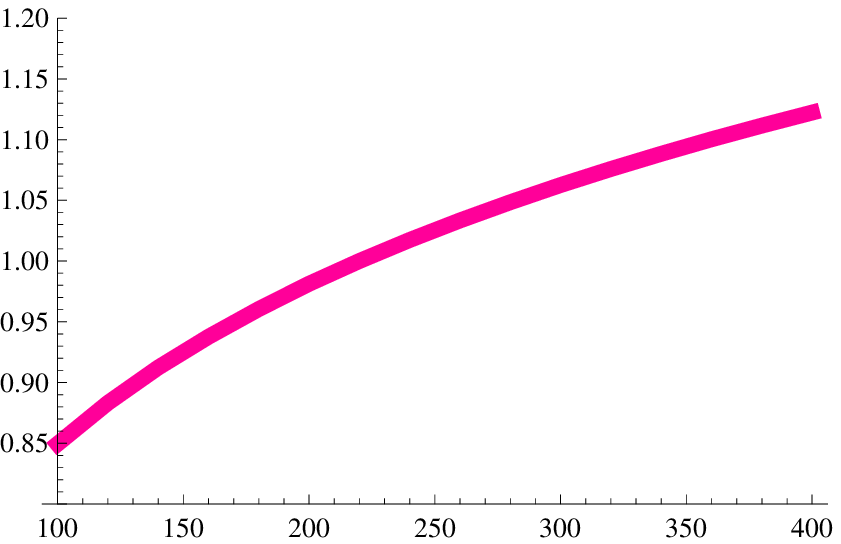}h)\\
\includegraphics[scale=0.6]{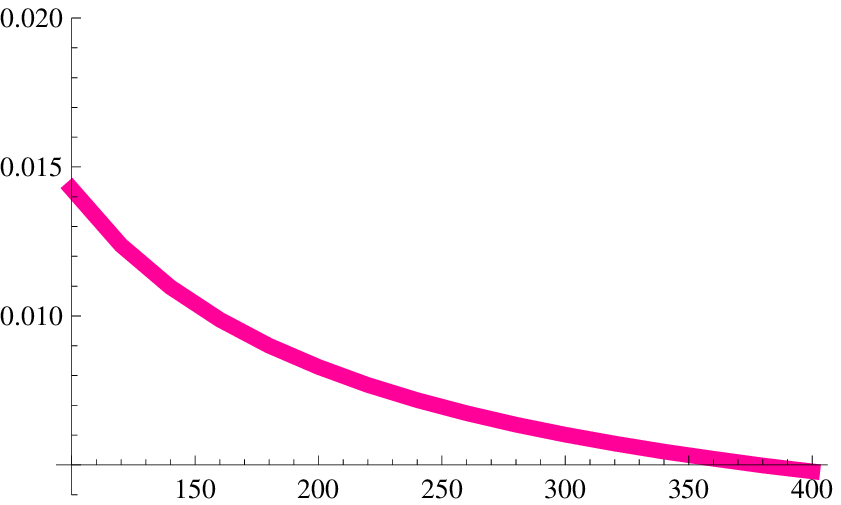}$\,\,\,$d)
\qquad\qquad
\includegraphics[scale=0.6]{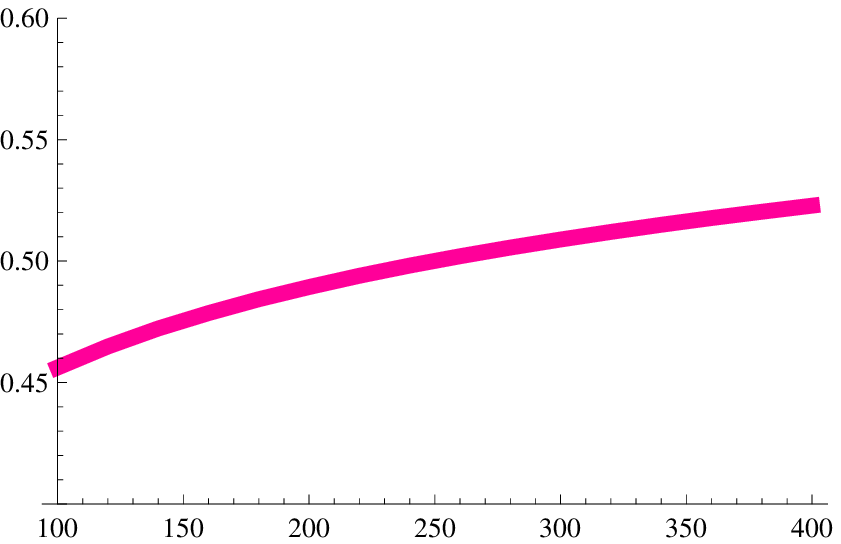}i)\\
\includegraphics[scale=0.6]{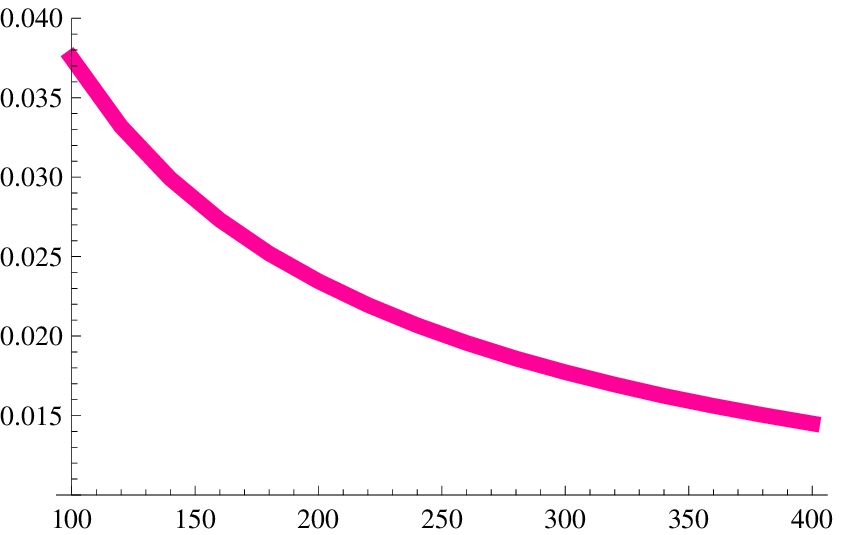}$\,\,\,$e)
\qquad\qquad
\includegraphics[scale=0.6]{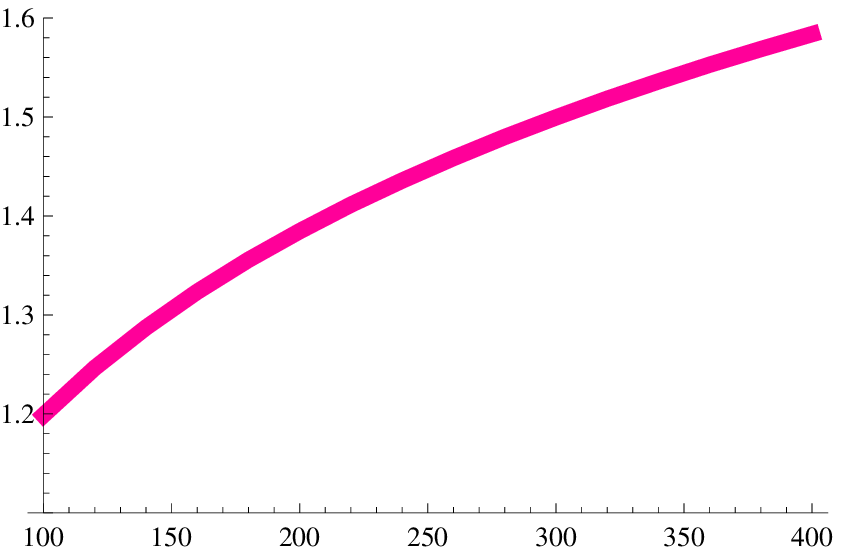}j)
\caption{$a)-e)$\,\,Graphs show the effect of the different
truncations in the calculation of the fermionic determinants
$Z_3^\psi Z_3^{\beta\gamma}$; $f)-j)$\,\,show dependence on $N$ of
the total $\mathcal{Z}^{osc}_3$. } \label{graph2}
\end{center}
\end{figure}

To perform numerical calculations of $\mathcal{Z}^{osc}_3$ we use
$N\times N$ matrix approximations for all matrices in the RHS's of
(\ref{Z}). The detailed results of the numerical calculations of
$\mathcal{Z}^{osc}_3$ for $N=100,120,...,400$ are presented in
Fig.$\ref{graph1}$ and Fig.$\ref{graph2}$. Namely,
\begin{itemize}
\item
in Fig.\ref{graph1}~a) $Z^{X}_3Z^{b,c}_3$ for $D=10$ is presented; a
fit of the form $a+bN^c$ gives
$$Z^{X}_3Z^{b,c}_3=-1.141+0.566 N^{0.879};$$
\item to see the difference with the previous
calculations performed by Fuchs and Kroyter \cite{KF} in
Fig.\ref{graph1}~b) we present the same results for $D=26$;
\item in Fig.\ref{graph2} a) and f)
$Z^{\psi}_3Z^{\beta,\gamma}_3$ and $\mathcal{Z}^{osc}_3$ are
presented, respectively; a fit of the form $a+b(\log N)^c$ gives
$$\mathcal{Z}^{osc}_{3,f} =0.11+0.36(\log N)^{0.76},$$
(here we use the representation for the Neumann matrices in the
superghost sector from \cite{GJ})
\item  in Fig.\ref{graph2}~b) -- $\,\,$e) and $\,$g) -- $\,$j)
$Z^{\psi}_3Z^{\beta,\gamma}_3$ and $\mathcal{Z}^{osc}_3$ for $D=10$
and different ways of truncations of the fermionic matrices
(\ref{V-matrices}) are presented. These ambiguities appear in the
factorized form (\ref{V-matrices}) due to a lack  of commutativity
for finite dimensional approximations of  matrices $F$ and
$\tilde{F}$;  fits for these data are
$$\mathcal{Z}^{osc}_{3,g}=0.176+0.02(\log N)^{0.17};$$
$$\mathcal{Z}^{osc}_{3,h}=0.16+0.09(\log N)^{1.27}; $$
$$\mathcal{Z}^{osc}_{3,i}= 0.24+0.04(\log N)^{1.06};$$
$$\mathcal{Z}^{osc}_{3,j}=0.41+0.07 (\log N)^{1.52};$$
here subscripts refer to the corresponding figures.
\end{itemize}


We see that the factorized form of the fermionic matrices produces
less singular answers, but still the coefficients
$\mathcal{Z}^{osc}_3$
 diverge logarithmically.
\newpage
$~$
\newpage
\section{Conclusion}
We have checked descent relations $ \langle V_2\mid
V_1\rangle=\mathcal{Z}_2 \langle V_1|$ and $ \langle V_3\mid
V_1\rangle=\mathcal{Z}_3 \langle V_2|$ for NS sector of SSFT. We
have performed calculations  in the usual oscillator basis and in
the  $\kappa$ basis. We have found  unexpected situation with the
normalization factor $\mathcal{Z}_3$.
\begin{itemize}
\item First, our  calculations show that starting from vertices subject
to overlap relations we as a rule get nontrivial $\mathcal{Z}_3$
\item Second, different schemes of calculations gives different $\mathcal{Z}_3$
and therefore $\mathcal{Z}_3$ has no universal meaning.
\end{itemize}

In both  schemes of calculation used in the paper the vertex is
factorized into the vertices of the bosonic and fermionic matter and
their ghosts and we checked the descent relations separately for
each vertex. These vertices  have produced four coefficients $Z_3$.
In the total descent relation we got the product of all these
factors $\mathcal{Z}_3=Z^X Z^{bc} Z^{\psi} Z^{\beta\gamma}$.

We start our calculation in $\kappa$-basis.
The regularized spectral density (\ref{ro}) has the divergent and
finite parts.
Due to the special tuning of the regularization
the divergent parts of regularized spectral
densities are the same for arbitrary conformal weights. Just due to this choice we have got the nonsingular  $\mathcal{Z}_3^{\kappa}$.
If one uses regularized
spectral densities different from  given in (\ref{ro})
one gets a different answer. For example, working with the same regularized spectral densities
for fermionic and bosonic sectors we would get an other finite part for $\mathcal{Z}_3^{\kappa}$.
Generally speaking using other regularization scheme one even cannot guarantee that
$\mathcal{Z}_3$ will be finite. If we didn't use the regularization of
the inner product, we would probably get the divergent factor $\mathcal{Z}_3$ in the
$\kappa$-basis.
The level truncation method demonstrates also an appearance of divergencies.
Namely, performing
 calculation of the descent relation in the discrete basis
we got that the factor $\mathcal{Z}^{osc}_3$ logarithmically
diverges. One can say that in the oscillation scheme the
regularization using truncations of the infinite Neumann matrices by
$N\times N$ matrices appears to be unlucky and brings divergencies
for $\mathcal{Z}^{osc}_3$. It can happen that exist a special
truncation method that provide a finite $Z_3$ in SSFT. Note that the
truncation method used in \cite{KF2} for the bosonic string give the
$\mathcal{Z}^{x,bc}_3=1$.

Therefore, the different methods and schemes of the
calculations produce the different regularizations.
 There are several papers \cite{Belov2,KF,AGMR,KF2,FKM} in which the
factor $\mathcal{Z}_3=Z^XZ^{bc}$ was calculated in the bosonic string.
Different methods of
calculations have been used and
 different factors $\mathcal{Z}_3$ have been obtained. This is in agreement with our general discussion.

The same technique can be used to get descent relations in the alternative
formulation of SSFT \cite{Berkovich}.

\section*{Acknowledgements}

We would like to thank D.~Belov and A.Pogrebkov for useful
discussions. We also are very grateful to our referee for
constructive critics of the preliminary version of the paper.

The work is supported in part by RFBR grant 05-01-00758 and Russian
President's grant NSh-672.2006.1. The work of I.A. is supported in
part by INTAS grant 03-51-6346. The work of D.R. is supported in
part by Dynasty foundation.
\appendix
\section{Neumann matrices $V^{ab}$ and $I$}
\setcounter{equation}{0}
 The Neumann matrices $V^{ab}$ and $I$ for the vertices
 $\langle V_3^\psi|$ and $|V_1^\psi\rangle$ have the form
\cite{AG,MS}: \begin{subequations}
\label{V-matrices} \begin{gather}
V^{aa}=\frac{F\tilde{F}}{(1-F)(2+F)},\\ V^{aa+1}=\frac{\tilde{F}+i
C(1-F)}{(1-F)(2+F)},\\ V^{aa-1}=\frac{\tilde{F}-i C(1-F)}{(1-F)(2+F)},\\
I=\frac{\tilde{F}}{1-F}=-\frac{1+F}{\tilde{F}}. \end{gather}
\end{subequations}
Here
the matrices $F$, $\tF$ and $C$ take the form  \cite{GJ}
\begin{eqnarray}
F_{rs}&=&-\frac{2}{\pi}\frac{i^{r-s}}{r+s},\quad r=s\text{ mod}(2),\nonumber\\
\tilde{F}_{rs}&=&\frac{2}{\pi}\frac{i^{r+s}}{s-r},\quad r=s+1\text{ mod}(2),\\
C_{rs}&=&(-1)^r\delta_{rs}.\nonumber
\end{eqnarray}
with the following properties
\begin{gather} \label{F-property1} F^{2}-\tilde{F}^{2}=1,\quad [F,\tilde{F}]=0,\\
 CFC=-F,\quad F^{T}=F,\quad
C\tilde{F}C=\tilde{F},\quad \tilde{F}^{T}=-\tilde{F}\nonumber. \end{gather}

At first we evaluate the  matrix $1-IV^{11}$:
\begin{equation}
1-IV^{11}=1+\frac{1+F}{\tF}\frac{F\tF}{(1-F)(2+F)}=\frac{2}{(1-F)(2+F)}.
\end{equation}
Let us consider the diagonal elements of matrix $R^{pq}$:
\begin{eqnarray*}
R^{pp}&=&V^{pp}+V^{p1}\frac{1}{1-IV^{11}}IV^{1p}=\\
&=&\frac{F\tF}{(1-F)(2+F)}+ \frac{\tF\mp
iC(1-F)}{(1-F)(2+F)}\frac{(1-F)(2+F)}{2}\frac{\tF}{1-F}\frac{\tF\pm
iC(1-F)}{(1-F)(2+F)}=\\ &=&\frac{F\tF}{(1-F)(2+F)}+\frac12 (\tF\mp
iC(1-F))^2\frac{\tF}{(1-F)^2(2+F)}\\ &=&\frac{F\tF}{(1-F)(2+F)}+\frac12
(\tF^2-(1-F)^2)\frac{\tF}{(1-F)^2(2+F)}=\nonumber\\
&=&\frac{F\tF}{(1-F)(2+F)}-\frac{(1-F)F\tF}{(1-F)^2(2+F)}=0.
\end{eqnarray*} Next we
evaluate the non-diagonal elements of $R^{pq}$:
\begin{eqnarray*}
R^{pp\pm1}&=&V^{pp\pm1}+V^{p1}\frac{1}{1-IV^{11}}IV^{1p\pm1}=\\ &=&\frac{\tF\pm
iC(1-F)}{(1-F)(2+F)}+\frac{\tF\mp
iC(1-F)}{(1-F)(2+F)}\frac{(1-F)(2+F)}{2}\frac{\tF}{(1-F)}\frac{\tF\mp
iC(1-F)}{(1-F)(2+F)}=\\ &=&\frac{\tF\pm iC(1-F)}{(1-F)(2+F)}-\frac{(1\mp i\tF
C)\tF}{(1-F)(2+F)}=\pm\frac{iC(1-F-\tF^2)}{(1-F)(2+F)}=\pm iC.
\end{eqnarray*}

\section{Neumann matrices $K^{ab}$ and $\tI$}

 The Neumann matrices $K^{ab}$ and $\tI$ for the vertices
 $\langle V_3^{\beta\gamma}|$ and $|V_1^{\beta\gamma}\rangle$ have the form
\begin{subequations} \label{zero-vert} \begin{gather}
K^{aa}=\frac{F\tilde{F}}{(1+F)(2-F)},\\ K^{aa+1}=\frac{-\tilde{F}-i
C(1+F)}{(1+F)(2-F)},\\ K^{aa-1}=\frac{-\tilde{F}+i C(1+F)}{(1+F)(2-F)},\\
\tI_{rs}=-\frac{\tilde{F}}{1+F}=\frac{1-F}{\tilde{F}}. \end{gather}
\end{subequations} Using the representation of the Neumann matrices (\ref{zero-vert})
we can evaluate analytically $U^{pq}$ like in the case of the fermionic matter. At
first we evaluate the inverse matrix $1-\tI K^{11}$: \begin{equation} 1-\tI
K^{11}=1-\frac{1-F}{\tF}\frac{F\tF}{(1+F)(2-F)}=\frac{2}{(1+F)(2-F)}. \end{equation}
Let us consider the diagonal elements of matrix the $U^{pq}$: \begin{eqnarray*}
U^{pp}&=&K^{pp}+K^{p1}\frac{1}{1-\tI K^{11}}\tI K^{1p}=\\ &=&
\frac{F\tF}{(1+F)(2-F)}+\frac12 (-\tF\pm
iC(1+F))^2\frac{\tF}{(1+F)^2(2-F)}=\nonumber\\
&=&\frac{F\tF}{(1+F)(2-F)}-\frac{(1+F)F\tF}{(1+F)^2(2-F)}=0. \end{eqnarray*} Next we
evaluate the non-diagonal elements of $U^{pq}$: \begin{eqnarray*}
U^{pp\pm1}&=&K^{pp\pm1}+K^{p1}\frac{1}{1-\tI K^{11}}\tI K^{1p\pm1}=\\
&=&\frac{-\tF\mp iC(1+F)}{(1+F)(2-F)}+\frac{(1\mp i\tF
C)\tF}{(1+F)(2-F)}=\mp\frac{iC(1+F-\tF^2)}{(1+F)(2-F)}=\mp iC. \end{eqnarray*}

 \end{document}